\documentclass[preprint,aps]{revtex4}
\usepackage[dvipdfmx]{graphicx}
\def\vc#1{\mbox{\boldmath $#1$}}
\usepackage{color}

\begin{document}
 
\title{{Isoscalar monopole excitations in $^{16}$O:~$\alpha$-cluster states at low energy and mean-field-type states at higher energy}}
 
\author{
	{Taiichi Yamada},
	{Yasuro Funaki$^{1}$},
	{Takayuki Myo$^{2}$},
	{Hisashi Horiuchi$^{3}$}, \\
	{Kiyomi Ikeda$^{1}$},
	{Gerd R\"opke$^{4}$},
	{Peter Schuck$^{5}$},
	{Akihiro Tohsaki$^{6}$}
}
 
\affiliation{Laboratory of Physics, Kanto Gakuin University, Yokohama 236-8501, Japan,}
  
\affiliation{$^{1}$Nishina Center for Accelerator-based Science, The Institute of Physical and Chemical Research (RIKEN), Wako 351-0098, Japan,}

\affiliation{$^{2}$General Education, Faculty of Engineering, Osaka Institute of Technology, Osaka 535-8585, Japan,}
 
\affiliation{$^{3}$Research Center for Nuclear Physics (RCNP), Osaka University, Ibaraki, Osaka 567-0047, Japan, and International Institute for Advanced Studies, Kizugawa, Kyoto 619-0225, Japan,}
   
\affiliation{$^{4}$Institut f\"ur Physik, Universit\"at Rostock, D-18051 Rostock, Germany,}
 
\affiliation{$^{5}$Institut de Physique Nucl\'eaire, CNRS, UMR 8608, Orsay, F-91406, France, and Universit\'e Paris-Sud, Orsay, F-91505, France, and Labratoire de Physique et Mod\'elisation des Milieux Condens\'es, CNRS et Universit\'e Joseph Fourier, 25 Av. des Martyrs, BP 166, F-38042 Grenoble Cedex 9, France,}
 
\affiliation{$^{6}$Research Center for Nuclear Physics (RCNP), Osaka University, Ibaraki, Osaka 567-0047, Japan}

\date{\today}
 
\begin{abstract}
Isoscalar monopole strength function in $^{16}$O up to $E_{x}\simeq40$ MeV is discussed. We found that the fine structures at the low energy region up to $E_{x} \simeq 16$ MeV in the experimental monopole strength function obtained by the $^{16}$O$(\alpha,\alpha^{\prime})$ reaction can be rather satisfactorily reproduced within the framework of the $4\alpha$ cluster model, while the gross three bump structures observed at the higher energy region ($16 \lesssim  E_{x} \lesssim 40$ MeV) look likely to be approximately reconciled by the mean-field calculations such as RPA and QRPA. In this paper, it is emphasized that two different types of monopole excitations exist in $^{16}$O;~one is the monopole excitation to cluster states which is dominant in the lower energy part ($E_{x} \lesssim 16$ MeV), and the other is the monopole excitation of the mean-field type such as one-particle one-hole ($1p1h$) which {is attributed} mainly to the higher energy part ($16 \lesssim E_{x} \lesssim 40$ MeV).  It is found that this character of the monopole excitations originates from the fact that the ground state of $^{16}$O with the dominant doubly closed shell structure has a duality of the mean-field-type {as well as} $\alpha$-clustering {character}. This dual nature of the ground state seems to be a common feature in light nuclei.
\\
\\
{PACS numbers: 23.20.-g, 21.60.Gx, 27.20.+n
}\\
\end{abstract}
 
\maketitle
 
\section{Introduction}\label{introduction}
 
Isoscalar monopole excitation in nuclei provides important information on its underlying structure. In the collective liquid drop model, the isoscalar {giant monopole} resonance (ISGMR), which has been established in medium and heavy nuclei~\cite{youngblood77}, corresponds to a breathing mode of the nucleus arising due to in-phase oscillations of the proton and neutron fluids. In heavy nuclei, the ISGMR is observed as a single peak in the $\alpha$ inelastic scattering cross sections at small angles, and its excitation energy follows an empirical formula $E_{x}\simeq80A^{-1/3}$~MeV, which is directly related to the compressibility of nuclear matter. A lot of work has been done to extract experimentally the nuclear compressibility by comparing {it} with microscopic calculations, for example, using the random phase approximation (RPA). Recently the isoscalar monopole distributions in $^{90}$Zr, $^{116}$Sn, $^{144}$Sm, and $^{208}$Pb were measured with greater precision than previously~\cite{youngblood99}. {The} results indicated that the compressibility of nuclear matter is $K_{\rm nm}=231\pm5$ MeV.
 
It is interesting to study what happens {for the} ISGMR in lighter nuclei. When the nuclear {masses decrease} from medium nuclei to light nuclei, the surface-energy correction {becomes} more important and the excitation energy of ISGMR should become lower compared with the empirical formula, indicating {a} lower nuclear compressibility~\cite{pandharipande70,blaizot76,lebrun80}. A lot of theoretical work has been so far devoted to the study of ISGMR in light nuclei, for example, {within the RPA framework}~\cite{blaizot76,drozdz80,paar06,papakonstantinou07,papakonstantinou09,ma97} and others~\cite{furuta10}.
 
The RPA calculations with {the} non-relativistic framework were performed in $^{16}$O, $^{40}$Ca, $^{90}$Zr, and $^{208}$Pb~\cite{blaizot76,drozdz80,paar06,papakonstantinou07,papakonstantinou09}. According to the results with the Gogny force and Skyrme {forces} etc.~\cite{blaizot76}, {it was found} that 1)~the isoscalar monopole strength in $^{16}$O spreads out over some energy {region} of $20\lesssim E_{x} \lesssim 40$ MeV with its centroid energy being $E_{x}=22\sim29$ MeV, the value of which depends on the $NN$ interactions employed, 2)~the monopole strength becomes more and more concentrated in a single peak as the nucleus becomes heavier, 3)~the percentage of energy-weighted sum rule carried by the resonances increases with the mass number of the nucleus, and 4)~the nuclear compressibility becomes smaller as the nucleus becomes lighter. On the other hand, Ma et al.~investigated the isoscalar monopole modes in $^{16}$O, $^{40}$Ca, $^{90}$Zr, and $^{208}$Pb using the relativistic RPA (RRPA) method~\cite{ma97}. They found that when going from heavy to lighter nuclei, {the} single-peak structure of ISGMR in $^{208}$Pb {changes} to a peak with several small humps in $^{40}$Ca and eventually the monopole {strength spreads} out {widely} to form a couple of peaks in $^{16}$O. Hence, {it was} pointed out that it becomes difficult, in a nucleus like $^{40}$Ca, to define theoretically the energy and width of the ISGMR.
 
\begin{figure}[t]
\begin{center}
\includegraphics[scale=0.6]{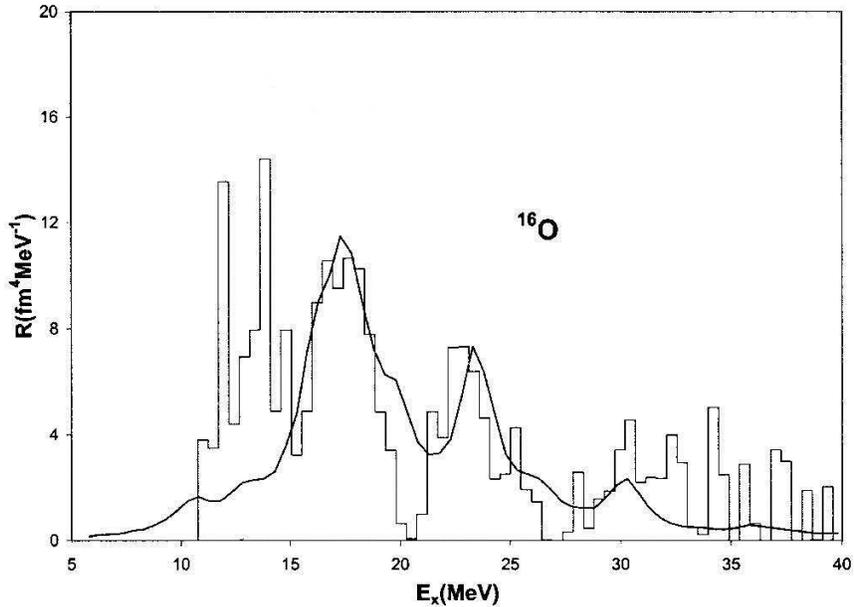}
\caption{
Experimental isoscalar monopole strength function of $^{16}$O~\cite{lui01} is shown by the histogram. The experimental data below $E_x\approx 10$~MeV are absent because of an energy cut in the experimental condition. The real line is the calculated result by the relativistic RPA calculation~\cite{ma97} multiplied by $0.25$ and shifted down in energy by $4.2$ MeV. This figure is taken from Ref.~\cite{lui01}.
}
\label{fig:1}
\end{center}
\end{figure}
 
The experimental isoscalar monopole strengths with a great precision were recently provided in $^{12}$C, $^{16}$O and $^{24}$Mg up to $E_{x}\simeq 50$~MeV, using inelastic scattering {of} $\alpha$ particles, by the Texas group~\cite{lui01}. They found that the isoscalar monopole strength in light nuclei {does} not concentrate on a single peak and the monopole strength {spreads} out in several regions {of energies}. The histogram in Fig.~\ref{fig:1} shows {the} experimental isoscalar monopole strength function in $^{16}$O~\cite{lui01}. {It is} compared with the RRPA calculation by Ma~et al.~\cite{ma97}. {It was} found that the centroid in the RRPA response function is at $25.3$~MeV, which is higher than the experimental data ($E_{x}=21.13\pm0.49$ MeV). In order to match their calculation to the experimental centroid, the calculated strength function was shifted down in energy by $4.2$~MeV and {furthermore they} normalized it to approximately 30\% of the isoscalar energy weighted sum rule (EWSR) by multiplying the RRPA curve by a factor of $0.25$~\cite{lui01}. Then, the {normalized and shifted} curve and the experimental result are in moderately good agreement {with each other with respect to} the shape of the gross three-peak structure. However, their calculation failed to reproduce the $0^{+}$ states found {in} the low energy region ($5 \lesssim E_{x} \lesssim 16$~MeV), in particular, at $E_{x}=6.05$, $12.05$ and $14.1$ MeV observed in inelastic $\alpha$ scattering and electron scattering etc.~\cite{lui01,ajzenberg93}. According to the $^{16}$O($e,e'$) experiments~\cite{ajzenberg93}, the three states are excited rather strongly by the $(e,e^{\prime})$ reaction, and their monopole matrix elements are $3.55\pm0.21$, $4.03\pm0.09$, and $3.3\pm0.7$~fm$^2$, respectively, comparable to the single-particle monopole strength~\cite{yamada08}. The total percentage of the energy weighted strength to the isoscalar monopole EWSR (energy weighted sum rule) for these three $0^+$ states amounts to be as large as over $15~\%$~\cite{ajzenberg93,yamada08}.
 
\begin{figure}[t]
\begin{center}
\includegraphics[scale=1.5]{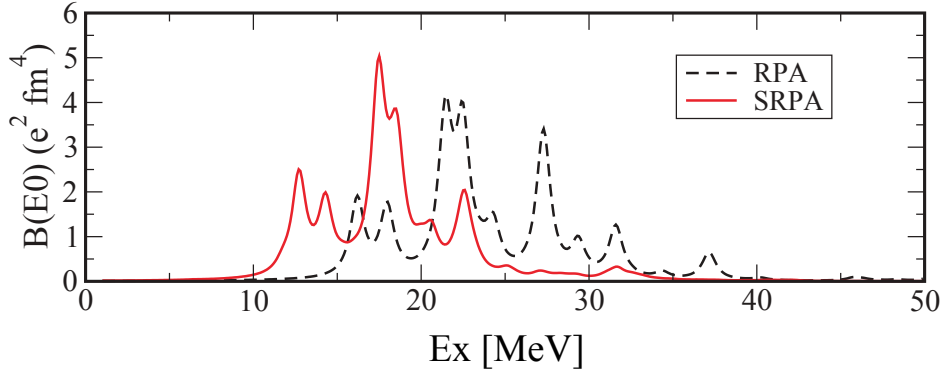}
\caption{
(Color online) RPA [dashed (black) lines] and SRPA [full (red) lines] for the isoscalar monopole strength distributions of $^{16}$O.
This figure is taken from Ref.~\cite{gambacurta10}.
}
\label{fig:srpa_monopole}
\end{center}
\end{figure}
 
In the nonrelativistic calculation for $^{16}$O~\cite{blaizot76} a significant discrepancy is also revealed as compared with the experimental data, in particular, in the low energy region ($5 \lesssim E_{x} \lesssim 16$~MeV), although the gross structures at the higher energy region ($E_{x} \gtrsim 20$ MeV) in the RPA calculations are in rather good agreement with the experimental data. This discrepancy in the low-energy region can also be seen in Fig.~\ref{fig:srpa_monopole} obtained by the recent second random-phase approximation (SRPA) calculations with a Skyrme force for $^{16}$O~\cite{gambacurta10}, in which the coupling between $1p1h$ and $2p2h$ as well as between $2p2h$ configurations among themselves are fully taken into account. In particular, their calculation fails to reproduce the monopole transition strength to the $0^+_2$ state at $E_x=6.05$~MeV observed by the $^{16}$O$(e,e')$ experiment. Thus, the monopole strengths in the lower energy region ($5 \lesssim E_{x} \lesssim 16$~MeV) are likely to be out of scope in the mean field theory. These results mean that the monopole strength function of $^{16}$O is not fully understood in the mean-field theory at the present stage, and other degree of freedoms beyond the mean field should be taken into account.
 
The $0^+_2$ and $0^+_3$ levels of $^{16}$O including its ground state, together with their monopole strengths, have in the past nicely been reproduced with a semimicroscopic cluster model, i.e.~the $\alpha$+$^{12}$C orthogonality condition model (OCM)~\cite{suzuki76}. The OCM is an approximation of the resonating group method (RGM)~\cite{saito68}. Many successful applications of OCM are reported in Ref.~\cite{ptp_supple_68}. The $\alpha$+$^{12}$C OCM calculation as well as the $\alpha$+$^{12}$C generator-coordinate-method one~\cite{libert80} demonstrates that the $0^+_2$ state at $E_x=6.05$~MeV and the $0^+_3$ state at $E_x=12.05$~MeV have $\alpha$+$^{12}$C structures, where the $\alpha$ particle orbits around the $^{12}$C($0^+$) core in an $S$ wave and around the $^{12}$C($2^+$) core in a $D$ wave, respectively. The 14.1-MeV $0^+$ state, however, could not be explained by the $\alpha$+$^{12}$C model calculations~\cite{suzuki76,libert80}.
 
Recently the structure study of $^{16}$O has made a great advance up to $E_x\simeq 16$~MeV around the $4\alpha$ disintegration threshold. The six lowest $0^+$ states of $^{16}$O, up to $E_{x} \simeq 16$~MeV, including the ground state, have for first time been reproduced very well with the $4\alpha$ OCM~\cite{funaki08}. The $4\alpha$ OCM shares {68~\%}of the energy weighted sum rule value of the isoscalar monopole transition of $^{16}$O, while the $\alpha$+$^{12}$C OCM does {31~\%}, as will be discussed below. Thus, it is interesting to investigate whether the $4\alpha$ OCM can reproduce the experimental isoscalar monopole strength function in the low energy region up to $E_{x}\simeq 16$ MeV in $^{16}$O, a region which is difficult to be treated in the mean-field theory. As will be discussed below, the five excited $0^{+}$ states of $^{16}$O up to $E_{x} \simeq 16$~MeV have $\alpha$-cluster structures~\cite{funaki08,suzuki76,libert80,ptp_supple_68}.
 
The purpose of the present paper {is} two {fold}:~first is to show that the isoscalar monopole strength function calculated with the $4\alpha$ OCM is in good correspondence to the experimental one in the low energy region up to $E_{x}\simeq 16$ MeV shown in Fig.~\ref{fig:1}, and {the} second is to emphasize {two features} in the isoscalar monopole excitation of $^{16}$O, i.e.~that the monopole excitation to cluster states is dominant in the lower energy part ($E_{x} \lesssim 16$ MeV) of the monopole strength function, whereas the monopole excitation of the $1p1h$-type {contributes} to the higher energy region ($16 \lesssim E_{x} \lesssim 40$ MeV). We will show that {the two features arise} from the {fact that} the ground state of $^{16}$O originally possesses {a} dual nature allowing $\alpha$-type {excitations} as well as $1p1h$-type {ones}, as will be discussed below. In this paper, {a} shell model calculation with the model space of $0s$-, $0p$-, $0d1s$-, and $0f1p$-shells for $^{16}$O is also performed to investigate the extent to which the shell model works for describing the low-lying $0^+$ states.
 
In Sec.~\ref{formulation}, the monopole excitation function with the $4\alpha$ OCM is formulated after a brief explanation of the $4\alpha$ OCM framework together with the shell-model framework for $^{16}$O. Results and discussions are devoted to Sec.~\ref{sec:results_discussion}, together with the energy weighted sum rule of the isoscalar monopole transition. Finally we present a summary in Sec.~\ref{sec:summary}.
 
\section{Formulation}\label{formulation}
 
First we formulate the isoscalar monopole strength function within the framework of the $4\alpha$ OCM. Then, the formulation of the shell model analysis is presented for $^{16}$O within the model space of $0s$, $0p$, $0d1s$, and $0f1p$ shells.
 
\subsection{Monopole strength function}\label{subsec:formulation_S(E)}
 
The strength function $S(E)$ of the monopole excitation from the $^{16}$O ground state $0^+_1$ is defined with use of the isoscalar monopole operator $\mathcal{O}=\sum_{i=1}^{16}  (\vc{r}_i - \vc{R}_{\rm cm} )^2$ as {follows},
\begin{eqnarray}
S(E) = \sum_{n}\delta(E-E_{n}) | {\langle  0^{+}_{n} | \sum_{i=1}^{16}  (\vc{r}_i - \vc{R}_{\rm cm} )^2 | 0^{+}_1 \rangle } |^{2}
\end{eqnarray}
where $\vc{r}_i$ ($i=1\sim16$) are the coordinates of nucleons, $\vc{R}_{\rm cm}={\frac{1}{16}}\sum_{i=1}^{16}\vc{r}_{i}$ is the c.o.m.~coordinate of $^{16}$O, and $E_{n}$ denotes the excitation energy of the $0^{+}_{n}$ state of $^{16}$O. On the other hand, the response function for the transition operator ${\cal O}$ is defined as
\begin{eqnarray}
R(E) =  {\langle  0^{+}_{1} | \frac{{\cal O}^{\dagger}{\cal O}}{E-H+i\epsilon}  | 0^{+}_{1} \rangle },
\end{eqnarray}
with $\epsilon$ representing an infinitesimal positive number. Then, $R(E)$ is related to $S(E)$ through
\begin{eqnarray}
S(E) = -\frac{1}{\pi}{\mathcal Im}\left[ R(E) \right] = \sum_{n} \delta(E-E_n) {|{\langle 0^+_n | {\cal O} | 0^+_1 \rangle}|}^{2}.
\end{eqnarray}
When the state $| 0^{+}_{n} \rangle$ is a {resonance} state with the complex energy $E_{n}-i\Gamma_{n}/2$, the strength function is expressed as
\begin{eqnarray}
&&S(E) = -\frac{1}{\pi}{\mathcal Im}\left[ R(E) \right]  = \frac{1}{\pi} \sum_{n} \frac{\Gamma_{n}/2}{(E-E_{n})^{2}+(\Gamma_{n}/2)^2} | \mathcal{M}({0^{+}_{n}}-{0^{+}_{1}}) |^{2}, \label{eq:S(E)}\\
&&\mathcal{M}({0^{+}_{n}}-{0^{+}_{1}}) = {\langle {0^{+}_{n}} | \sum_{i=1}^{16} (\vc{r}_i - \vc{R}_{\rm cm} )^2 | {0^{+}_{1}} \rangle},\label{eq:me_e0}
\end{eqnarray}
where $\Gamma_{n}$ represents the width of the $0^{+}_{n}$ state. The isoscalar monopole transition matrix element, $\mathcal{M}({0^{+}_{n}}-{0^{+}_{1}})$, has a {relation} with the ${\rm E0}$ transition matrix element $M({\rm E0},{0^{+}_{n}}-{0^{+}_{1}})$ for the $0^+_1$ and $0^+_n$ states with the total isospin $T=0$ as {follows},
\begin{eqnarray}
{M({\rm E0},{0^{+}_{n}}-{0^{+}_{1}})} &\equiv & {\langle  0^{+}_{n} | \sum_{i=1}^{16} \frac{1+\tau_{3i}}{2}  (\vc{r}_i - \vc{R}_{\rm cm} )^2 | 0^{+}_1 \rangle } \nonumber \\
                                           &=& {\frac{1}{2}} {\mathcal{M}({0^{+}_{n}}-{0^{+}_{1}})}.
\label{eq:me0_general}
\end{eqnarray}
The energy weighted sum rule (EWSR) of the isoscalar monopole transition~\cite{yamada08} reads
\begin{eqnarray}
&&\sum_{n} (E_n-E_1) {\left| \mathcal{M}({0^{+}_{n}}-{0^{+}_{1}}) \right|^{2}} = \frac{2\hbar^2}{m} \times 16 \times R^{2},\label{eq:EWSR_nucleon}\\
&&R^{2} = \frac{1}{16} {\langle  0^{+}_{1} | \sum_{i=1}^{16}  (\vc{r}_i - \vc{R}_{\rm cm} )^2 | 0^{+}_1 \rangle }, \label{eq:rms_gs_16O}
\end{eqnarray}
where $R$ and $m$ represent the r.m.s radius of the ground state and nucleon mass, respectively. Here, we assume that the $NN$ interaction has no velocity dependence. 
{Employing the experimental charge radius of $^{16}$O ($R_c=2.70$~fm~\cite{ajzenberg93}), the value of $R$ in Eq.~(\ref{eq:rms_gs_16O}) is estimated to be $2.58$~fm, in which the effects of the charge radius of proton (${\langle r^2 \rangle}_{\rm proton}=0.8791^2$~fm$^{2}$) and that of neutron (${\langle r^2 \rangle}_{\rm neutron}=-0.1149$~fm$^2$)~\cite{ajzenberg93} are subtracted from the charge radius of $^{16}$O ($R_c$): $R = \sqrt{{R_c}^2 - {\langle r^2 \rangle}_{\rm proton} - {\langle r^2 \rangle}_{\rm neutron}} = 2.58$~fm. Then, the total EWSR value, $\frac{2\hbar^2}{m} \times 16 \times R^{2}${,} is $8.83 \times 10^3$~fm$^{4}\cdot$MeV}.  
 
It is instructive to see a characteristic feature of the isoscalar monopole operator in Eq.~(\ref{eq:me_e0}), which can be decomposed into two parts, internal parts and relative parts, with respect to $4$ $\alpha$ clusters in $^{16}$O (as well as $\alpha$ and $^{12}$C clusters in $^{16}$O). Since the operator in Eq.~(\ref{eq:me_e0}) has a quadratic form with respect to the coordinates of nucleons, the following interesting identities are realized,
\begin{eqnarray}
\sum_{i=1}^{16} (\vc{r}_i - \vc{R}_{\rm cm} )^2 &=& {\sum_{k=1}^{4} \sum_{i=1}^{4} ( \vc{r}_{i+4(k-1)} - \vc{R}_{\alpha_k} )^2} + \sum_{k=1}^{4} {4(\vc{R}_{\alpha_k}-\vc{R}_{\rm cm})^2}, \label{eq:e0_4alpha}\\
                                                       &=& {\sum_{i=1}^{4} ( \vc{r}_{i} - \vc{R}_{\alpha} )^2} + {\sum_{i=5}^{16} ( \vc{r}_{i} - \vc{R}_{\rm C} )^2} + 3{\vc{\xi}_3}^2,\label{eq:e0_alpha_12C}
\end{eqnarray}
where $\vc{R}_{\alpha_k}=(1/4)\sum_{i=1}^{4}\vc{r}_{i+4(k-1)}$ is the c.o.m.~coordinate of the $k$-th $\alpha$ cluster, and $\vc{\xi}_j$ ($j=1\sim3)$ are Jacobi coordinates 
with respect to the c.o.m.~coordinates of {$4$~$\alpha$ clusters} ($\vc{R}_{\alpha_k}$, $k=1\sim4$):~$\vc{\xi}_{1}=\vc{R}_{2}-\vc{R}_{1}$, $\vc{\xi}_{2}=\vc{R}_{3}-(\vc{R}_1+\vc{R}_2)/2$, and $\vc{\xi}_3=\vc{R}_4-(\vc{R}_1+\vc{R}_2+\vc{R}_3)/3$. {In Eq.~(\ref{eq:e0_alpha_12C}), $\vc{R}_{\alpha} = (1/4)\sum_{i=1}^{4}\vc{r}_{i}$ and  $\vc{R}_{\rm C} = (1/12)\sum_{i=5}^{16}\vc{r}_{i}$ stand for the c.o.m.~coordinates of $\alpha$ and $^{12}$C clusters, respectively.} Here we should {recall the} useful identity of $\sum_{k=1}^{4} {4(\vc{R}_{\alpha_k}-\vc{R}_{\rm cm})^2} = {\sum_{j=1}^{3} \mu_{k}{\vc{\xi}_{j}}^{2}}$ in Eq.~(\ref{eq:e0_4alpha}), {where $\mu_{j}$ ($\mu_1=2$, $\mu_2=8/3$, and $\mu_3=3$) correspond to the reduced masses with respect to the Jacobi coordinates $\vc{\xi}_j$}.
 
Equation~(\ref{eq:e0_4alpha}) {shows} that the monopole operator consists of two parts:~1) {internal} parts [first term in the right hand of Eq.~(\ref{eq:e0_4alpha})] {composed of} the {internal} coordinates of each $\alpha$-cluster, and 2) relative parts [second term in the right hand of Eq.~(\ref{eq:e0_4alpha})] acting {on} the relative motions of the $4\alpha$ clusters with respect to the c.o.m. of $^{16}$O. On the other hand, Equation~(\ref{eq:e0_alpha_12C}) shows that the monopole operator can also be decomposed into two other parts:~1) two {internal} parts, {i.e.}~first and second terms in the right hand of Eq.~(\ref{eq:e0_alpha_12C}), {composed of} the {internal} coordinates of the $\alpha$ and $^{12}$C clusters, respectively, and 2)~relative part acting {on} the relative motion between the $\alpha$ and $^{12}$C clusters. The fact that the isoscalar monopole operator consists of the two parts, the {internal} part and the relative part, {plays} an important role in {the monopole excitation of $^{16}$O}, {see Sec.~\ref{sec:results_discussion}}.
 
\subsection{$4\alpha$ OCM}\label{sub:4a_ocm}
 
The total wave function {$\tilde{\Psi}(J^{\pi})$} of the $4\alpha$ system with total angular momentum $J^{\pi}$ in the OCM framework is {expressed by the product of the internal wave functions of $\alpha$ clusters $\phi(\alpha)$ and the relative wave function $\Psi(J^{\pi})$ among the $4\alpha$ clusters
\begin{eqnarray}
\tilde{\Psi}(J^{\pi}) = \Psi(J^{\pi}) \phi(\alpha_1) \phi(\alpha_2) \phi(\alpha_3) \phi(\alpha_4).
\end{eqnarray}
}
The relative wave function $\Psi(J^{\pi})$ is expanded in terms of Gaussian basis functions as follows,
\begin{eqnarray}
&&\Psi(J^{\pi})=\sum_{c, \nu} A_{c}(\nu)\Phi_{c}(\nu), \label{eq:total_wf}\\
&&\Phi_{c}(\nu) ={\cal \widehat S} \Big[[\varphi_{l_1} (\vc{\xi}_1,\nu_1)\varphi_{l_2}(\vc{\xi}_2,\nu_2)]_{l_{12}}  \varphi_{l_3} (\vc{\xi}_3,\nu_3) \Big]_{J}, \label{eq:total_wf_basis}\\
&&{\langle u_{F} | \Psi(J^{\pi}) \rangle} = 0,\label{eq:PF}
\end{eqnarray}
where $\vc{\xi}_1$, $\vc{\xi}_2$ and $\vc{\xi}_3$ are the Jacobi coordinates describing internal motions of the $4\alpha$ system. ${\cal \widehat S}$ stands for the symmetrization operator acting on all $\alpha$ particles obeying Bose statistics. $\nu$ denotes the set of size parameters $\nu_1, \nu_2$ and $\nu_3$ of the normalized Gaussian function, $\varphi_{l}(\vc{\xi},\nu_i)=N_{l,\nu_i}\xi^l\exp{(-\nu_i \xi^2)} Y_{l m}(\hat{\vc \xi})$, and $c$ the set of relative orbital angular momentum channels $[[l_1,l_2]_{l_{12}},l_3]_J$ depending on either of  the coordinate type of $K$ or $H$~{\cite{funaki08,GEM}}, where $l_1$, $l_2$ and $l_3$ are the orbital angular momenta with respect to the corresponding Jacobi coordinates. Equation~(\ref{eq:PF}) represents the orthogonality condition that the total wave function (\ref{eq:total_wf}) should be orthogonal to the Pauli-forbidden states of the $4\alpha$ system, $u_F$'s, which are constructed from Pauli forbidden states between two $\alpha$-particles in $0S$, $0D$ and $1S$ states~\cite{horiuchi_77}. The ground state with {the dominant} shell-model-like configuration $(0s)^{4}(0p)^{12}$ can be described properly in the present $4\alpha$ OCM framework{, as discussed below}.
 
The $4\alpha$ Hamiltonian for {$\Psi(J^{\pi})$} is given as follows:
\begin{eqnarray}
\mathcal{H} &=&\sum_{i}^{4}T_i - T_{\rm cm}+ \sum_{i<j}^4 \Big[ V_{2\alpha}^{({\rm N})}(i,j)+V^{({\rm  C})}_{2\alpha}(i,j) \Big]  \nonumber \\
&+&\sum_{i<j<k}^4 V_{3\alpha}(i,j,k)+ V_{4\alpha}(1,2,3,4), \label{eq:hamil}
\end{eqnarray}
where $T_i$, $V_{2\alpha}^{({\rm N})}(i,j)$, $V_{2\alpha}^{({\rm C})}(i,j)$, $V_{3\alpha}(i,j,k)$ and $V_{4\alpha}(1,2,3,4)$ stand for the operators of kinetic energy for the $i$-th $\alpha$ particle, two-body, Coulomb, three-body and four-body forces between $\alpha$ particles, respectively. The center-of-mass kinetic energy $T_{\rm cm}$ is subtracted from the Hamiltonian. The effective $\alpha$-$\alpha$ interaction $V_{2\alpha}^{\rm (N)}$ is constructed by the folding procedure from an effective two-nucleon force. Here we take the Modified Hasegawa-Nagata (MHN) force~\cite{mhn} as the effective $NN$ force, {which is constructed based on the $G$-matrix theory}. It is noted that the folded $\alpha$-$\alpha$ potential reproduces the $\alpha$-$\alpha$ scattering phase shifts and energies of the $^8$Be ground state and of the Hoyle state. The three-body force is phenomenologically introduced so as to fit the ground state energy of the $^{12}$C with the framework of the $3\alpha$ OCM. The same force parameter set as used in Ref.~\cite{yamada05_3a_ocm} is adopted in the present calculation. In addition, the phenomenological four-body force is adjusted to the ground state energy of $^{16}$O. The three-body and four-body forces are short-range, and, hence, they only act in compact configurations. The coefficients $A_{c}(\nu)$ in Eq.~(\ref{eq:total_wf}) are determined according to the Rayleigh-Ritz variational principle.
 
The isoscalar monopole matrix element is evaluated as follows:
\begin{eqnarray}
\mathcal{M}^{\rm OCM}({0^{+}_{n}}-{0^{+}_{1}}) 
&=& {{\langle \tilde{\Psi}({0^{+}_{n}}) |  {\sum_{i=1}^{16} (\vc{r}_{i}-\vc{R}_{\rm cm})^2}  | \tilde{\Psi}({0^{+}_{1}}) \rangle},} \label{eq:me_e0_total_ocm}\\
&=& {\langle \Psi({0^{+}_{n}}) |  {\sum_{k=1}^{4} {4(\vc{R}_{\alpha_k}-\vc{R}_{\rm cm})^2}}  | \Psi({0^{+}_{1}}) \rangle} + {{16\times {R(\alpha)^{2}} \delta_{n1}}},\label{eq:me_e0_ocm}\\
&=& {\langle \Psi({0^{+}_{n}}) |  2{\vc{\xi}_1}^2 + \frac{8}{3}{\vc{\xi}_2}^2 +  3{\vc{\xi}_3}^2  | \Psi({0^{+}_{1}}) \rangle} + {{16\times {R(\alpha)^{2}} \delta_{n1}}}   ,\label{eq:me_e0_ocm_jacobi}
\end{eqnarray}
where $\Psi$ is the $4\alpha$ OCM wave function in Eq.~(\ref{eq:total_wf}). In Eqs.~(\ref{eq:me_e0_total_ocm})$\sim$(\ref{eq:me_e0_ocm_jacobi}) we used the relation,
\begin{eqnarray}
{\langle \phi(\alpha_1) \phi(\alpha_2) \phi(\alpha_3) \phi(\alpha_4) |  {\sum_{k=1}^{4} \sum_{i=1}^{4} ( \vc{r}_{i+4(k-1)} - \vc{R}_{\alpha_k} )^2}  | \phi(\alpha_1) \phi(\alpha_2) \phi(\alpha_3) \phi(\alpha_4) \rangle} = 16 \times {R(\alpha)^{2}},
\end{eqnarray}
where {$R(\alpha)$} is the r.m.s.~radius of $\alpha$ particle, ${R(\alpha)}=\sqrt{\frac{1}{4}\langle \sum_{i=1}^{4} (\vc{r}_{i} - \vc{R}_{\alpha})^{2} \rangle_{\alpha}}$~\cite{yamada05_3a_ocm,funaki08}. It is {important} to study the EWSR of the isoscalar monopole transition within the framework of the $4\alpha$ OCM. We call it the OCM-EWSR, and its definition reads
\begin{eqnarray}
&&\sum_{n} (E_n-E_1) {\left| \mathcal{M}^{\rm OCM}({0^{+}_{n}}-{0^{+}_{1}}) \right|^{2}} \nonumber \\
&&\hspace{20mm} = \frac{1}{2} {\langle \Psi({0^+_1}) | [\mathcal{O}_{\rm OCM},[\mathcal{H},\mathcal{O}_{\rm OCM}] | \Psi({0^+_1})\rangle}, \\
&&\hspace{20mm} = \frac{2\hbar^2}{m} {\langle {\Psi({0^+_1})} | {\sum_{k=1}^{4} {4(\vc{R}_{\alpha_k}-\vc{R}_{\rm cm})^2}} | {\Psi({0^+_1})} \rangle},\\
&&\hspace{20mm} = \frac{2\hbar^2}{m} \times 16 \times ( R^2 - {R(\alpha)^{2}} ) \label{eq:EWSR_cluster}
\end{eqnarray}
where $\mathcal{H}$ is given in Eq.~(\ref{eq:hamil}) {and $\mathcal{O}_{\rm OCM} =  {\sum_{k=1}^{4} {4(\vc{R}_{\alpha_k}-\vc{R}_{\rm cm})^2}}$} [see Eq.~(\ref{eq:me_e0_ocm})], and $R$ denotes the r.m.s.~radius of the $^{16}$O ground state given in Eq.~(\ref{eq:rms_gs_16O}). It is noted that the $4\alpha$ OCM can describe the shell-model-like structure of the $^{16}$O ground state, as shown later. Then, the ratio of the OCM-EWSR to the total EWSR in Eq.~(\ref{eq:EWSR_nucleon}) is
\begin{equation}
\frac{{\rm OCM\mbox{\small -}EWSR}}{\rm total~EWSR} = 1 - \left(\frac{{R(\alpha)}}{R}\right)^{2} 
= { 1 - \left(\frac{1.47}{2.58}\right)^{2}= 0.68.} \label{eq:ratio_total_OCM_EWSR}
\end{equation}  
{Here we use $R(\alpha)=1.47$~fm and $R=2.58$~fm, which are estimated from the experimental charge radii ($1.68$~fm and $2.70$~fm, respectively~\cite{ajzenberg93}) with subtracting the effects of the charge radius of proton and that of neutron from them, the method of which is the same as that shown in previous section.}
This result means that the $4\alpha$ OCM framework shares about {$70~\%$} of the total EWSR value {(the OCM-EWSR is also discussed in Appendix)}. This is one of the important reasons that the $4\alpha$ OCM works rather well in reproducing the isoscalar monopole transitions in the low-energy region of $^{16}$O as shown later. 
 
In the present paper, the energies $E_n$ and isoscalar monopole matrix elements $\mathcal{M}$ in Eq.~(\ref{eq:S(E)}) are obtained by the $4\alpha$ OCM calculation. As for the widths $\Gamma_n$, we estimate the $\alpha$-decay widths with the $R$-matrix theory~\cite{lane58},
\begin{eqnarray}
&&\Gamma_L =2P_L(a) \gamma^2_L(a), \\
&&P_L(a)=\frac{ka}{F_L^2(ka)+G_L^2(ka)}, \\
&&\gamma^2_L(a)=\theta^2_L(a)\gamma^2_{\rm W}(a), \\
&&\gamma^2_{\rm W}(a)=\frac{3\hbar^2}{2\mu a^2},
\label{eq:gamma}
\end{eqnarray}
where $k$, $a$ and $\mu$ are the wave number of the {$\alpha$-$^{12}$C} relative motion, the channel radius, and the reduced mass, respectively, and $F_L$, $G_L$, and $P_L(a)$ are the regular and irregular Coulomb wave functions and the corresponding penetration factor, respectively. The reduced width of $\theta^2_L(a)$ is related to the reduced {width} amplitude or overlap amplitude ${\cal Y}_{L}$ as, $\theta^2_L(a)=\frac{a^3}{3}{\cal Y}_L^2(a)$, and the definition of ${\cal Y}_{L}$ is presented as
\begin{equation}
{\cal Y}_{L}(r)= \sqrt{\frac{4!}{3!1!}} \Big\langle \Big[ \frac{\delta(r^\prime-r)}{r^{\prime 2}} Y_{L}(\vc{\hat r}^\prime)\Phi_{L}(^{12}{\rm C}) \Big]_{0} \Big| \Psi(0^+_n) \Big\rangle.
\label{eq:rwa}
\end{equation}
Here, $\Phi_{L}(^{12}{\rm C})$ is the wave function of $^{12}$C, given by the $3\alpha$ OCM calculation~\cite{yamada05_3a_ocm}, and $r$ is the relative distance between the center-of-mass of $^{12}$C and the $\alpha$ particle. {The spectroscopic factor} of the $\alpha$+$^{12}$C($L$) channel {$S^2_L$} in the $0^+_n$ state of $^{16}$O, {defined as}
\begin{eqnarray}
{S^{2}_L} = \int_{0}^{\infty} {dr} \left[ r\mathcal{Y}_{L}(r) \right]^2,\label{eq:s2_factor}
\end{eqnarray}
is useful to analyze the obtained wave functions.
 
In the present study, we perform more careful analyses than the previous ones~\cite{funaki08}, in particular, for identifying the $0^+$ states around the $4\alpha$ threshold. The calculation of the resonant state in the bound state approximation is usually done by diagonalizing the Hamiltonian with use of a finite number of square-integrable basis wave functions. The positive-energy eigenstates obtained by the diagonalization are divided into resonant states and continuum states, and many methods for carrying out the division are proposed~\cite{kukulin_book89}. In the present study, a pseudopotential method is adopted to divide the resonant states and continuum states, as shown below.
 
Let's us first consider a repulsive pseudopotential $V$ that is added to the original Hamiltonian $H$, yielding
\begin{eqnarray}
H^{\prime}(\delta) = H + {\delta \times V},
\end{eqnarray}
where $\delta$ is a constant used to vary the strength of the pseudopotential. As increasing {into negative values} the constant $\delta$ from the physical value, $\delta=0$, the eigenenergy of this new Hamiltonian $H^{\prime}(\delta)$ decreases for any resonance state, which is eventually transformed into a bound state. On the contrary, continuum states show almost no change in their eigenvalues as $\delta$ increases into the negative region. In the present $4\alpha$ OCM framework, it is important to study the eigenenergies with changing the constant $\delta$ but with no change in the threshold energies of the $\alpha$+$^{12}$C and $4\alpha$ decay channels, even though we introduce the pseudopotential $V$. Here, we take the four-body potential $V_{4\alpha}$ in Eq.~(\ref{eq:hamil}) as the pseudopotential $V$, because the choice is convenient {for practical reasons in} the present numerical calculation. This pseudopotential method is simple but  helpful to identify the resonant states under the bound state approximation. As a result, we obtained almost the same results as the previous ones~\cite{funaki08}, as will be shown {below}.

\subsection{Shell model calculation}\label{sub:shell_model}
 
The shell model Hamiltonian of $^{16}$O adopted here is presented as follows:
\begin{eqnarray}
H = {\sum_{i=1}^{16}t_i} - T_{\rm cm} + \sum_{i<j=1}^{16}\left[ v^{\rm (C)}(i,j)+ v^{\rm (LS)}(i,j) \right],
\end{eqnarray}
where $t_i$ denotes the kinetic energy of the $i$-th nucleon, and {$v^{\rm (C)}$ ($v^{\rm (LS)}$)} represents the central ($LS$) force of the effective $NN$ interaction.
The c.o.m. kinetic energy $T_{\rm cm}$ is subtracted from the total kinetic energy. The model space adopted here {covers} all configurations of $1p1h$ and $2p2h$ within the  $0s$, $0p$, $0d1s$, and $0f1p$ shells. The spurious states of the c.o.m.~motion are eliminated with the Lawson's method~\cite{lawson80}.
 
In the present study, we take the Volkov No.~2 force~{\cite{volkov65}} and G3RS force~{\cite{tamagaki68}} for {$v^{\rm (C)}$ and $v^{\rm (LS)}$,} respectively.  The Majorana parameter ($M$) in the Volkov No.2 force{, the multiplying factor ($\zeta_0^{\rm LS}$) of the G3RS, and} the nucleon size parameter ($b$) are chosen so as to reproduce the total binding energies of the ground states of $^{16}$O and $^{15}$O, the $LS$-splitting between $3/2^{-}_1$ and $1/2^{-}_1$ in $^{15}$O, and the r.m.s.~radius of the ground state of $^{16}$O as well as possible. The following two parameter sets are adopted:~case A for $(M,\zeta_0^{\rm LS},b)=(0.665,1.7,1.70~{\rm fm})$ and case B for $(0.620,2.6,2.0~{\rm fm})$. The present shell-model code is based on the code in {Refs.~\cite{myo07,myo11}}.
 
\begin{figure}[t]
\begin{center}
\includegraphics[scale=0.6]{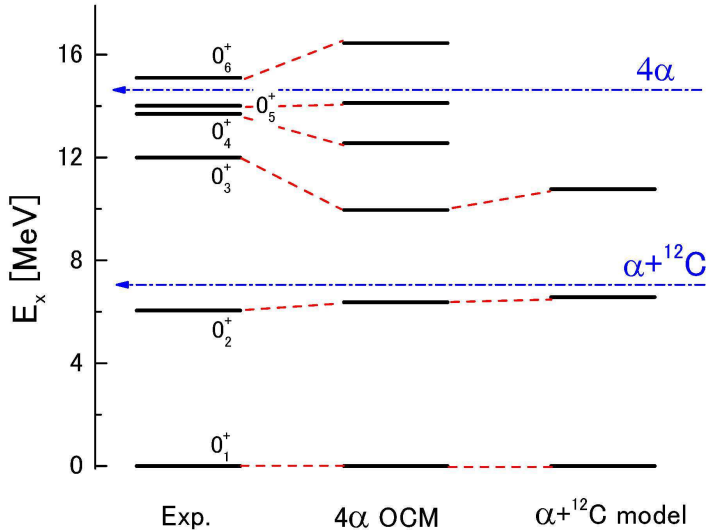}
\caption{
(Color online) Comparison of energy spectra among experiment, the $4\alpha$ OCM calculation~\cite{funaki08}, and the $\alpha$+$^{12}$C model calculation~\cite{suzuki76}, where the $\alpha$+$^{12}$C and $4\alpha$ thresholds are shown.  Experimental data are taken from Ref.~\cite{ajzenberg93} and from Ref.~\cite{wakasa07} for the $0_4^+$ state.
}
\label{fig:2}
\end{center}
\end{figure}
 
\section{Results and discussion}\label{sec:results_discussion}
 
\subsection{$4\alpha$ OCM calculation}\label{subsec:4a_OCM}
 
The energy levels of $0^+$ states in $^{16}$O obtained by the present $4\alpha$ OCM calculation are shown in Fig.~\ref{fig:2} and Table~\ref{tab:4a_ocm}.  One can make the one-to-one correspondence of the six lowest $0^+$ states observed up to $E_{x}\simeq 16$ MeV in the {$4\alpha$} OCM calculation. It is reminded that the $\alpha$+$^{12}$C {OCM} cluster model~\cite{suzuki76} can reproduce only the lowest three $0^+$ states. {We} obtained almost the same results as the previous {$4\alpha$ OCM calculation}~\cite{funaki08}. The six $0^+$ states have the following characteristic structures~\cite{funaki08}:~1)~the ground state ($0^{+}_{1}$) has {dominantly} a doubly-closed-shell structure, 2)~the $0_2^+$ state at $E_{x}=6.05$ MeV and the $0_3^+$ state at $E_{x}=12.05$ MeV have {mainly} $\alpha + ^{12}$C structures~\cite{horiuchi68} where the $\alpha$-particle orbits around the $^{12}$C$(0_1^+)$ core in an $S$-wave and around the $^{12}$C$(2_1^+)$ core in a $D$-wave, respectively, the results of which are consistent with the previous studies with $\alpha + ^{12}$C OCM~\cite{suzuki76} and the $\alpha+^{12}$C Generator Coordinate Method (GCM)~\cite{libert80}, 3)~the $0_4^+$ ($E_{x}=13.6$ MeV) and $0_5^+$ ($E_{x}=14.1$ MeV) states mainly have $\alpha + ^{12}$C$(0_1^+)$ structure with higher nodal behavior and $\alpha + ^{12}$C$(1^-)$ structure, respectively, where in the latter the $\alpha$ particle moves around the $^{12}$C$(1^-)$ core (corresponding to the first $1^-$ state at $E_x=10.84$ MeV having an intermediate structure between the shell-model-like structure and cluster structure~\cite{ptp_supple_68}) in $P$-orbit, and 4)~the $0^{+}_{6}$ state at $15.1$ MeV is a strong candidate of the $4\alpha$ condensate, $(0S)^{4}_{\alpha}$, with the probability of $61~\%$.
 
\begin{table}[t]
\begin{center}
\caption{Excitation energies ($E_{x}$), {charge} r.m.s.~radii ({$R_c$}), ${\rm E0}$ transition matrix elements [$M({\rm E}0)$], and particle decay widths ($\Gamma$) of the $0^+$ states in $^{16}$O obtained by the $4\alpha$ OCM calculation and $\alpha$+$^{12}$C {OCM} model calculation~\cite{suzuki76}, together with the experimental data~\cite{ajzenberg93,wakasa07}. They are given in the unit of MeV, fm, fm$^2$, and MeV, respectively. The experimental monopole matrix elements are obtained by the $^{16}$O($e,e^{\prime}$) reaction~\cite{ajzenberg93}. $P^{\rm e.w.}$ represents the percentage of the energy weight strength to the isoscalar monopole EWSR [see Eq.~(\ref{eq:EWSR_nucleon})]. The finite size effects of $\alpha$ particle and $^{12}$C are taken into account in estimating {$R_c$} with the $4\alpha$ OCM and $\alpha$+$^{12}$C OCM (see Ref.~\cite{yamada05_3a_ocm} for details). }
\label{tab:4a_ocm}
\begin{tabular}{ccccccccccccccc}
\hline\hline
      & \multicolumn{4}{c}{$4\alpha$ OCM} &  & \multicolumn{3}{c}{$\alpha$+$^{12}$C {OCM}} & &  \multicolumn{5}{c}{Experiment} \\
           & \hspace{3mm}{$E_{x}$}\hspace{3mm} & \hspace{3mm}{{$R_c$}}\hspace{3mm} & \hspace{1mm}{$M({\rm E}0)$}\hspace{1mm} & {$\Gamma$} & \hspace*{3mm} &  \hspace{3mm}{$E_{x}$}\hspace{3mm} & \hspace{3mm}{{$R_c$}}\hspace{3mm} & \hspace{1mm}{$M({\rm E}0)$}\hspace{1mm} & \hspace*{3mm} & $E_{x}$ &\hspace{1mm}{$R_c$}\hspace{1mm} & \hspace{1mm}$M({\rm E}0)$\hspace{1mm} & $P^{\rm e.w.}$ &  {$\Gamma$} \\
\hline
\hspace{3mm}$0_1^+$\hspace{3mm} & $ 0.00 $   & $2.7$ &        &          &  & $0.00$ & $2.5$ &          & & $ 0.00 $ & {$2.70$} &                       &   &         \\
$0_2^+$                                 & $ 6.37 $   & $3.0$ & $3.9$ &          &  & $6.57$ & $2.9$ & $3.88$ & & $ 6.05 $ &          &   $3.55\pm 0.21$  & {$3.5\%$} & \\
$0_3^+$                                 & $ 9.96 $   & $3.1$ & $2.4$ &          &  & $10.77$ & $2.8$ & $3.50$ & & $ 12.05 $ &        & $4.03\pm 0.09$   & {$8.9\%$} &\\
$0_4^+$                                 & $ 12.56$   & $4.0$ & $2.4$ & {$0.60$} &  & $-$    & $-$   & $-$   & & $13.60$  &           &  no data            &          &$0.6$    \\
$0_5^+$                                 & $ 14.12$   & $3.1$ & $2.6$ & {$0.20$} &  & $-$    & $-$   & $-$   & & $14.01$ &           &  $3.3\pm0.7$      & {$6.9\%$} & $0.185$  \\
$0_6^+$                                 & $ 16.45$   & $5.6$ & $1.0$ & {$0.14$} &  & $-$    & $-$   & $-$   & & $15.10$  &          &  no data            &         & $0.166$ \\
\hline\hline
\end{tabular}
\end{center}
\end{table}
 
\begin{figure}[t]
\begin{center}
\includegraphics[width=40mm]{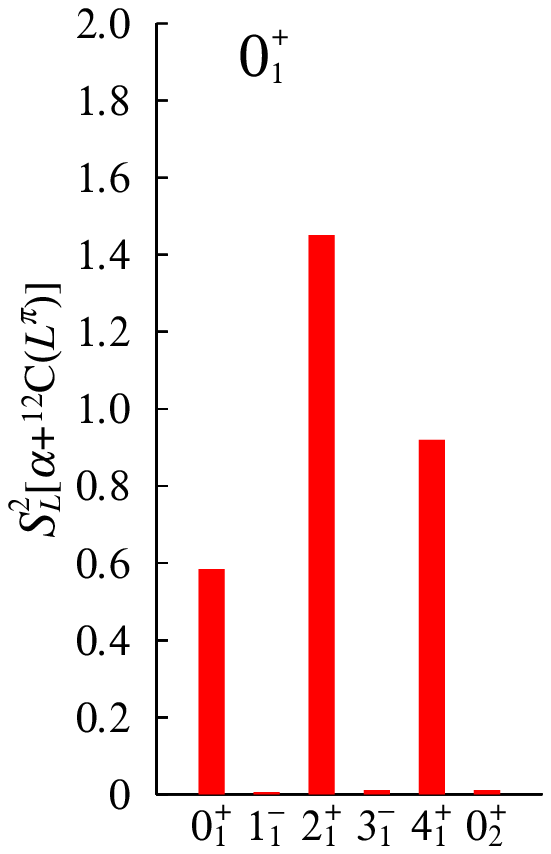}
\includegraphics[width=40mm]{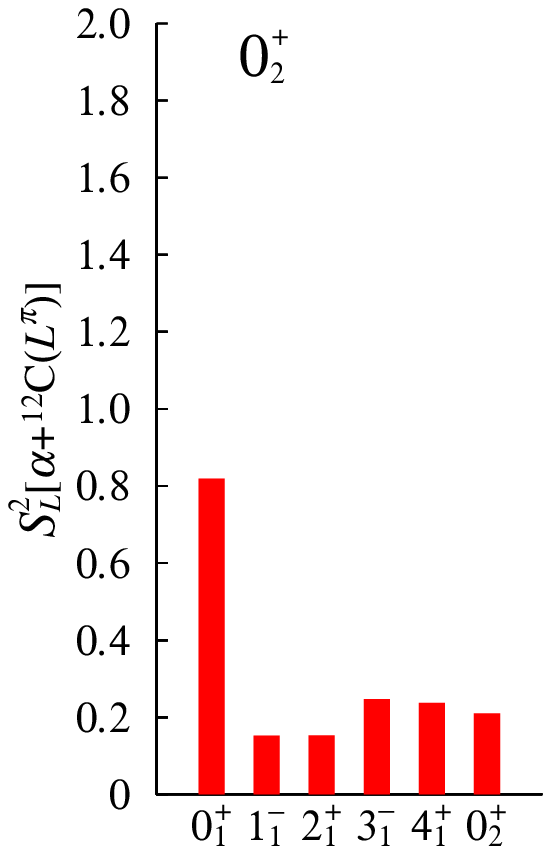}
\includegraphics[width=40mm]{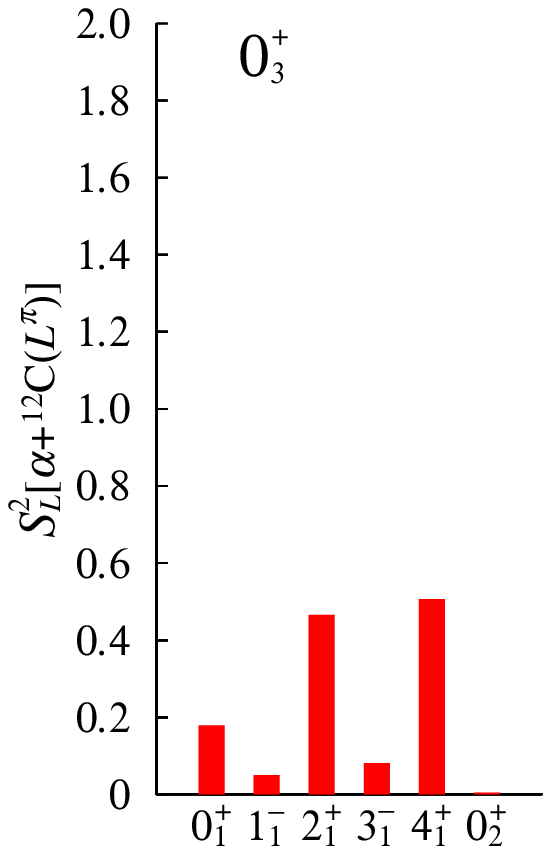}
\\ \vspace*{5mm}
\includegraphics[width=40mm]{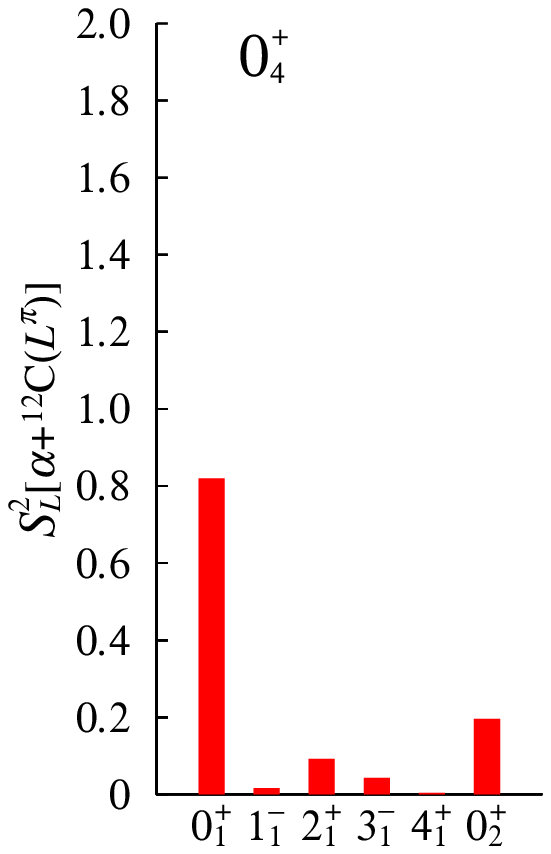}
\includegraphics[width=40mm]{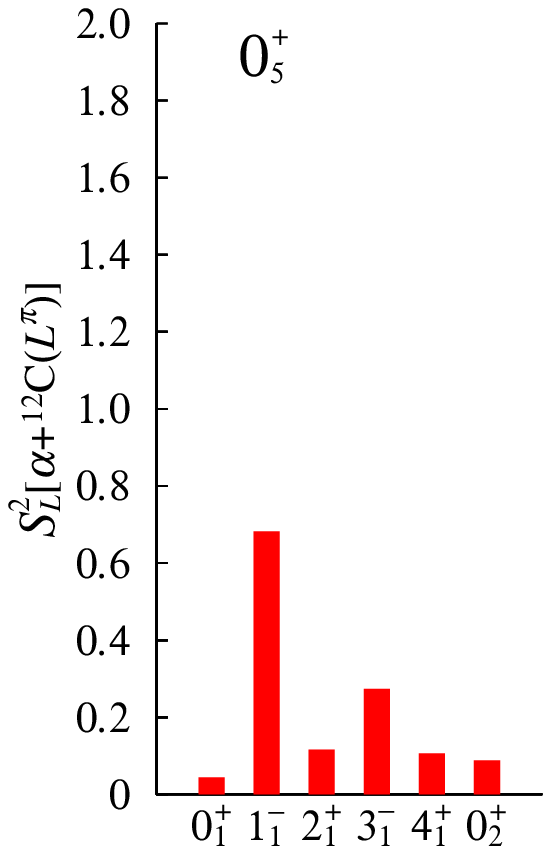}
\includegraphics[width=40mm]{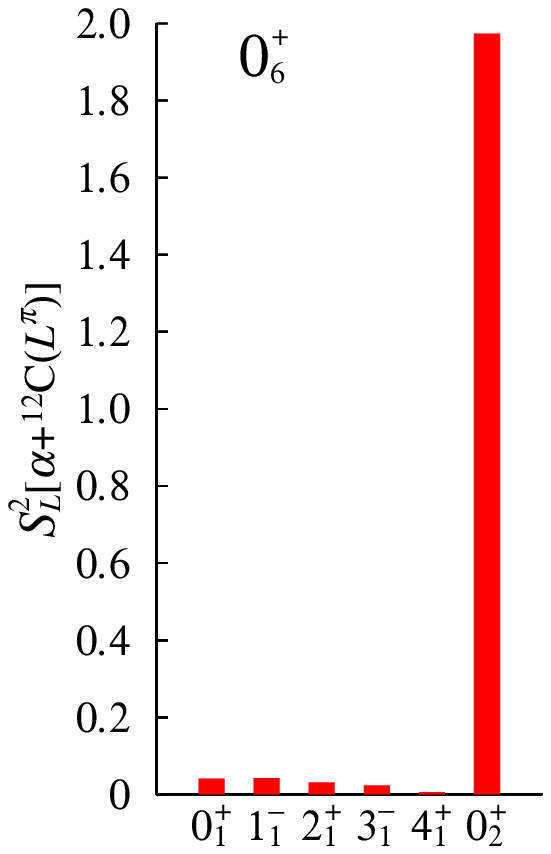}
\caption{{Spectroscopic factors $S^{2}_{L}$} of the $\alpha + ^{12}$C$(L^{\pi}_n)$ channels ($L^{\pi}_n=0^{+}_{1},1^{-}_{1},2^{+}_{1},3^{-}_{1},4^{+}_{1},0^{+}_{2}$) in the six $0^{+}$ sates of $^{16}$O {defined in Eq.~(\ref{eq:s2_factor})}.}
\label{fig:s2_factors_0+_states_16O}
\end{center}
\end{figure}
 
These characteristic features of the structures of the six $0^+$ states can be verified from the analysis of the {spectoscopic factors $S^{2}_{L}$} defined in Eq.~(\ref{eq:s2_factor}). The results are shown in Fig.~\ref{fig:s2_factors_0+_states_16O}. Since the ground state has a closed shell structure with the dominant component of SU(3)$(\lambda,\mu)=(0,0)$~\cite{elliott58}, the values of {the spectroscopic factors $S^2_{L}$} for $0^+_1$ in Fig.~\ref{fig:s2_factors_0+_states_16O} can be explained by the SU(3) nature of the state. This SU(3) character was confirmed by the recent {no-core} shell model~\cite{dytrych07}. As mentioned above, the structures of the $0_2^+$ and $0_3^+$ states are well established as having the $\alpha + ^{12}$C$(0_1^+)$ and $\alpha + ^{12}$C$(2_1^+)$ cluster structures, respectively. These structures of the $0_2^+$ and $0_3^+$ states are confirmed by the $4\alpha$ OCM calculation. In fact, one sees in Fig.~\ref{fig:s2_factors_0+_states_16O} that the {$S^2$ factors} for the $\alpha + ^{12}$C$(0_1^+)$ and $\alpha + ^{12}$C$(2_1^+)$ channels are dominant in the $0_2^+$ and $0_3^+$ states, respectively. {In the $0^+_3$ state, however, the {$S^2$ factors} of the $\alpha$+$^{12}$C$(4_1^+)$ channel in the $4\alpha$ OCM calculation is rather large as compared with the result of the $\alpha$+$^{12}$C OCM one~\cite{suzuki76}. This is due to the following facts:~1)~The calculated excitation energy of the $^{12}$C$(4_1^+)$ state in the present $4\alpha$ OCM calculation is underestimated {by} $E_x \sim 8$~MeV, while {it is set to the experimental value ($E_x = 14.1$~MeV) in} the $\alpha$+$^{12}$C OCM {calculation}, and 2)~thus in the $4\alpha$ OCM calculation a coupled-channel effect of the $\alpha$+$^{12}$C$(4_1^+)$ channel with the $\alpha$+$^{12}$C$(0_1^+,2_1^+)$ channel is {reinforced} and consequently, the {$S^2$ factor} of the $\alpha$+$^{12}$C$(4_1^+)$ channel becomes larger. We expect that the $S^2$ factor of the $\alpha$+$^{12}$C$(4_1^+)$ channel will be smaller when the excitation energy of $^{12}$C$(4_1^+)$ is properly reproduced in the $4\alpha$ OCM calculation.}
 
Table~\ref{tab:4a_ocm} lists the ${\rm E0}$ transition matrix elements $M({\rm E0})$. The $M({\rm E0})$ value of the $0^+_2$ state is reproduced well, while that of the $0^+_3$ state underestimates the experimental result, and this trend is similar to the result of the $\alpha$+$^{12}$C {OCM} model~\cite{suzuki76}. On the other hand, the $0_4^+$ and $0_5^+$ states mainly have the $\alpha + ^{12}$C$(0_1^+)$ structure with higher nodal behavior and an $\alpha + ^{12}$C$(1^-)$ structure, respectively. The ${\rm E0}$ transition matrix element of the $0^+_5$ state is reproduced nicely within the experimental error (see Table~\ref{tab:4a_ocm}). In Table~\ref{tab:4a_ocm}, the largest r.m.s. radius is about 5 fm for the $0_6^+$ state, the {wave function} of which has a large overlap amplitude with the $\alpha + ^{12}$C$(0^+_2)$ channel (see Fig.~3 in Ref.~\cite{funaki08}). Hence the {$S^2$ factor of the $\alpha + ^{12}$C$(0^+_2)$ channel is dominant in the $0^+_6$ state} (see Fig.~\ref{fig:s2_factors_0+_states_16O}), whereas those in the other channels are much suppressed. This {dominance} of the {$S^2$ factor} of the $\alpha + ^{12}$C$(0^+_2)$ channel is one of the evidences for the $0^+_6$ state being the $4\alpha$-condensate-state, $(0S)^{4}_{\alpha}$, because the Hoyle state has the main configuration, $(0S)^{3}_{\alpha}$~\cite{tohsaki01,funaki03,yamada05_3a_ocm}, and the overlap amplitude of $\langle (0S)^3_{\alpha} | (0S)^4_{\alpha} \rangle $ becomes large.
 
As for the decay widths of the $0^+_4$ and $0^+_5$ states, the results are shown in Table~\ref{tab:4a_ocm}. The calculated width of the $0_4^+$ state is {$\sim 600$} keV, which is quite a bit larger than that found for the $0_5^+$ state {$\sim 200$} keV. Both are quantitatively consistent with the corresponding experimental data, $600$ keV and $185$ keV, respectively. On the other hand, the decay width of the $0^+_6$ state is very small, {$140$} keV, in reasonable agreement with the corresponding experimental value of {166} keV, indicating that this state is unusually long lived. We should note that our calculation consistently reproduces the ratio of the widths of the $0_4^+$, $0_5^+$, and $0_6^+$ states, i.e. about $6:3:2$, respectively (see Table~\ref{tab:4a_ocm}).
 
Comparing the energy levels of the six $0^+$ states with the experimental monopole response function of $^{16}$O shown in Fig.~\ref{fig:1}, one notices that the energy positions of the fine structures in the low energy region ($10 \lesssim E_{x} \lesssim 16$~MeV) of the experimental response function {seem} to be in good correspondence with the energy levels of $0^{+}_{3}$, $0^{+}_{4}$, $0^{+}_{5}$, and $0^{+}_{6}$ in Fig.~\ref{fig:2} and Table~\ref{tab:4a_ocm}. It {should be} noted that the peak corresponding to the $0^{+}_{2}$ state at $E_x=6.05$ MeV is not visible in Fig.~\ref{fig:1}, because of an energy cut in the {experimental} condition~\cite{lui01}. Thus, it is {important} to study the isoscalar monopole strength function within the framework of the $4\alpha$ OCM calculation.

\begin{figure}[t]
\begin{center}
\includegraphics[scale=0.5]{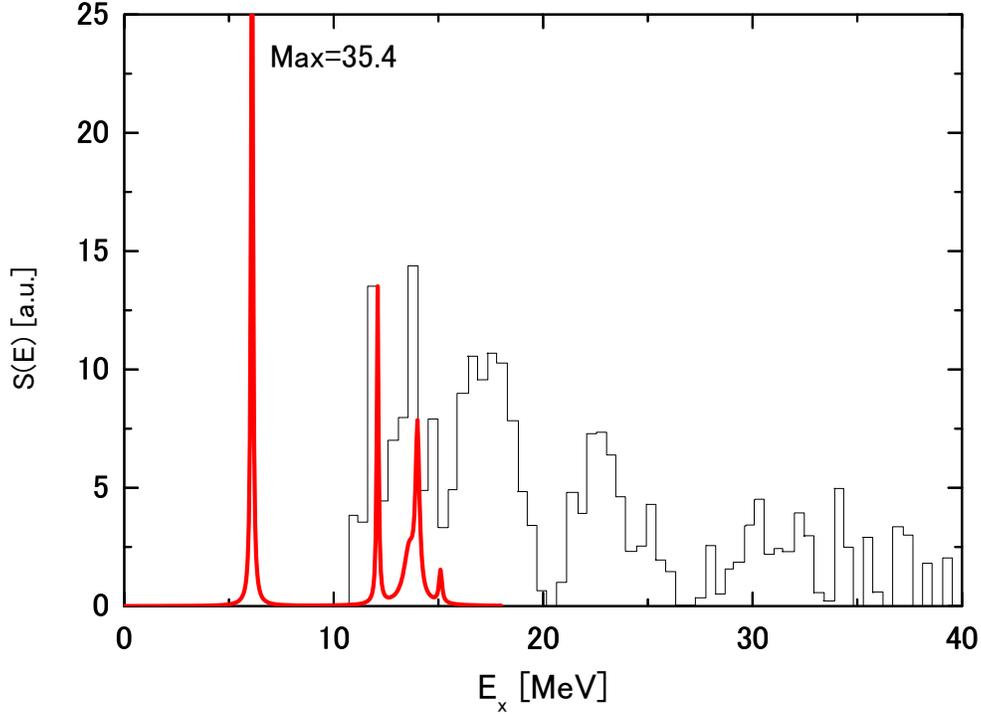}
\caption{
(Color online) Calculated isoscalar monopole strength functions of $^{16}$O (bold line) and experimental data (thin line: see Fig.~\ref{fig:1}~\cite{lui01}). 
Here we use the calculated monopole matrix elements and the calculated decay widths for the six $0^{+}$ states up to $E_{x} \simeq 16$ MeV obtained by the $4\alpha$ OCM calculation (see Table~\ref{tab:4a_ocm}), although the experimental excitation energies for the six $0^{+}$ states are employed (see text).
}
\label{fig:strength_fun_exp_cal}
\end{center}
\end{figure}
 
\subsection{Isoscalar monopole excitation function with the $4\alpha$ OCM calculation}
 
Figure~\ref{fig:strength_fun_exp_cal}(a) shows the calculated isoscalar monopole strength function of $^{16}$O defined in Eq.~(\ref{eq:S(E)}), where we use the calculated monopole matrix elements and the calculated decay widths for the six $0^{+}$ states up to $E_{x} \simeq 16$ MeV obtained by the $4\alpha$ OCM calculation (see Table~\ref{tab:4a_ocm}), {also} the experimental excitation energies for the six $0^{+}$ states are employed. {We take into account the experimental energy resolution of $50$ keV~\cite{lui01} for the width $\Gamma_n$ in Eq.~(\ref{eq:S(E)}) through $\Gamma_n=\sqrt{{\Gamma_n({\rm OCM})}^2+0.050^2}$, where ${\Gamma_n({\rm OCM})}$ denotes the calculated decay width of the $n$-th $0^+$ state of $^{16}$O given in Table~\ref{tab:4a_ocm}}. The calculated strength function is normalized so as to match the calculated strength of the 12.1-MeV peak to the experimental one. We can see a rather good correspondence with the experimental data. The fine structures in the calculated strength function, i.e.~one peak at $E_x = 12.1$~MeV (corresponding to the $0^+_3$ state), one shoulder-like peak at $E_x = 13.8$~MeV ($0^+_4$), two peaks at $E_x = 14.1$ MeV ($0^+_5$) and $15.1$ MeV ($0^+_6$), are well reproduced. As mentioned above, the fine structures in the energy region of $10 \lesssim E_x \lesssim 15$ MeV as well as the sharp peak at $E_x \simeq 6$ MeV (corresponding to the $0^+_2$ state) are difficult to be reproduced by any mean-field calculations~\cite{blaizot76,ma97,drozdz80,paar06,papakonstantinou07,papakonstantinou09,gambacurta10}, as {far} as the present authors know. The calculated values of $S(E)$ at each $0^+$ states ($0^{+}_{2} \sim 0^{+}_{6}$) in Fig.~\ref{fig:strength_fun_exp_cal} are approximately proportional to the squared values of the respective calculated monopole matrix elements.
 
It is instructive {and interesting} to discuss the mechanism of why the five $\alpha$ cluster states ($0^+_2$, $0^+_3$, $0^+_4$, $0^+_5$, and $0^+_6$) of $^{16}$O are excited relatively strongly from the ground state with the doubly-closed-shell-like structure~\cite{yamada08}. Their monopole matrix elements shown in Table~\ref{tab:4a_ocm} are comparable to the single particle strength ($\sim 5.4$~fm$^{2}$~\cite{yamada08}) and {share} about $20~\%$ of the total EWSR value. Since the mechanism is closely related {to} the property of the ground state of $^{16}$O as shown below, we first demonstrate its interesting properties with {the} use of the microscopic wave function and then discuss the monopole matrix elements in the OCM calculation.
 
The wave function of {the $^{16}$O ground state has dominantly} the doubly closed shell model configuration $(0s)^{4}(0p)^{12}$ with the nucleon size parameter $\nu=M\omega/2\hbar$ ($M$: nucleon mass), corresponding to the SU(3) $(\lambda,\mu)=(0,0)$ wave function~\cite{elliott58}. {This doubly closed shell model wave function} is mathematically equivalent to a single cluster model wave function of $\alpha$+$^{12}$C with the total harmonic oscillator quanta $Q=12$~\cite{yamada08},
\begin{eqnarray}
&&{\frac{1}{\sqrt{16!}} {\rm det} | (0s)^{4}(0p)^{12} |} \times [\phi_{\rm cm}(\vc{R}_{\rm cm})]^{-1} \label{eq:shell_wf_16O}\\
&& \hspace*{10mm} = N_{0} \sqrt{\frac{12!4!}{16!}} \mathcal{A} \left\{ \left[ u_{40}(\vc{\xi}_{3},3\nu)\phi_{L=0}({^{12}}{\rm C})\right]_{J=0} \phi(\alpha) \right\},\label{eq:su3_16O_0}\\
&& \hspace*{10mm} = N_{2} \sqrt{\frac{12!4!}{16!}} \mathcal{A} \left\{ \left[ u_{42}(\vc{\xi}_{3},3\nu)\phi_{L=2}({^{12}}{\rm C})\right]_{J=0} \phi(\alpha) \right\},\label{eq:su3_16O_2}\\
&& \phi_{\rm cm}(\vc{R}_{\rm cm}) = {\left(\frac{32\nu}{\pi}\right)^{3/4}} \exp(-16\nu{\vc{R}_{\rm cm}}^2),
\end{eqnarray}
where $\phi_{\rm cm}$ denotes the wave function of the c.o.m. motion of $^{16}$O, and $N_{0,2}$ are the normalization constants. $\phi(\alpha)$ and $\phi_{L}({^{12}}{\rm C})$ stand for the internal wave function of {the} $\alpha$ cluster with the $(0s)^{4}$ configuration and the internal wave function of $^{12}$C with the angular momentum of $L$ belonging to the SU(3) irreducible representation $(\lambda,\mu)=(0,4)$ for the $(0s)^{4}(0p)^{8}$ configuration. The relative wave {functions} between the $\alpha$ and $^{12}$C clusters in Eqs.~(\ref{eq:su3_16O_0}) and (\ref{eq:su3_16O_2}) {are} described by the harmonic oscillator wave function $u_{QLm}(\vc{\xi},\beta)=u_{QL}(\xi)Y_{Lm}(\hat{\vc{\xi}})$ with the node number $n=(Q-L)/2$ and $Q=4$. One can prove Eqs.~(\ref{eq:su3_16O_0}) and (\ref{eq:su3_16O_2}) with help of the Bayman-Bohr theorem~\cite{bayman58}. It should be reminded that the dominance of the doubly closed shell structure in the ground state of $^{16}$O is confirmed by the {no}-core shell model with realistic $NN$ forces~\cite{dytrych07}.
 
Equations~(\ref{eq:su3_16O_0}) and (\ref{eq:su3_16O_2}) mean that the doubly closed shell model wave function has an $\alpha$+$^{12}$C cluster degree of freedom. Since the monopole operator shown in Eq.~(\ref{eq:e0_alpha_12C}) has the relative part ($3{\vc{\xi}_3}^2$), this can populate $\alpha$+$^{12}$C cluster states by exciting the relative motion with respect to the relative coordinate ($\vc{\xi}_3$) between the $\alpha$ and $^{12}$C clusters. Since the $0^+_2$ and $0^+_3$ states of $^{16}$O have the dominant cluster structures $\alpha$+$^{12}$C($0^+_1$) and $\alpha$+$^{12}$C($2^+_1$), respectively, the monopole transition strengths between the ground state and the $\alpha$+$^{12}$C cluster states ($0^+_2$ and $0^+_3$) are enhanced {and} share $3~\%$ and $8~\%$ of the EWSR value, respectively~\cite{suzuki76,ptp_supple_68,yamada08}. The detailed explanation using the microscopic framework is given in Ref.~\cite{yamada08}.
 
In the $4\alpha$ OCM, the monopole matrix elements from the $^{16}$O ground state to the $\alpha$+$^{12}$C cluster states are evaluated with use of Eqs.~(\ref{eq:me_e0_ocm}) and (\ref{eq:me_e0_ocm_jacobi}). The validity of using the formulae is based on the following three facts found within the above-mentioned microscopic framework{~\cite{yamada08}}:~1)~the ground state of $^{16}$O is of the SU(3) $(\lambda,\mu)=(0,0)$ nature with the $\alpha$-clustering degree of freedom, 2) the monopole matrix elements come dominantly from the relative part of the monopole operator referring to the $\alpha$+$^{12}$C relative motion, $3{\vc{\xi}_3}^2$, 3)~the contribution from the other parts of the monopole operator becomes significantly smaller for the $\alpha$+$^{12}$C cluster states, {and 4)~the $\alpha$-cluster-type ground-state correlation significantly enhances the monopole strength compared with the case of the $^{16}$O ground state being the pure SU(3) $(\lambda,\mu)=(0,0)$ wave function}. These are certainly the reasons why the estimation of the monopole matrix element using Eqs.~(\ref{eq:me_e0_ocm}) and (\ref{eq:me_e0_ocm_jacobi}) in the {$4\alpha$ OCM} calculation gives a reasonable reproduction {for the experimental data}. In fact, the ground state of $^{16}$O obtained in the $4\alpha$ OCM has the dominant SU(3) $(\lambda,\mu)=(0,0)$ component with $Q=12$, and only the relative part with respect to the $\alpha$+$^{12}$C relative motion, $3{\vc{\xi}_3}^2$, in the monopole matrix element (\ref{eq:me_e0_ocm_jacobi}) gives the major contribution to the monopole matrix elements $M({\rm E0}; 0^+_1-0^+_{2,3})$ in the present {$4\alpha$} OCM calculation.
 
On the other hand, the reason why the $0^+_6$ state with the $4\alpha$-gas-like character has a relatively large monopole strength ($\sim 1$~fm$^{2}$) {can be also understood from} the property of the ground state of $^{16}$O. The doubly closed shell-model wave function in Eq.~(\ref{eq:shell_wf_16O}) is mathematically equivalent to a single $4\alpha$ cluster wave function with $Q=12$, according to the Bayman-Bohr theorem, as follows,
\begin{eqnarray}
&&{\frac{1}{\sqrt{16!}} {\rm det} | (0s)^{4}(0p)^{12} |} \times [\phi_{\rm cm}(\vc{R}_{\rm cm})]^{-1}  \nonumber \\
&& \hspace*{10mm} = \hat{N}_{0} \sqrt{\frac{4!4!4!4!}{16!}} \mathcal{A} \mbox{\LARGE \{} \left[ u_{40}(\vc{\xi}_{3},3\nu) \left[ u_{40}(\vc{\xi}_{2},\frac{8}{3}\nu) u_{40}(\vc{\xi}_{1},2\nu) \right]_{L=0} \right]_{J=0} \nonumber \\
&&\hspace*{30mm} \times \phi(\alpha_1) \phi(\alpha_2) \phi(\alpha_3) \phi(\alpha_4) \mbox{\LARGE \}},\label{eq:su3_16O_4a}
\end{eqnarray}
where $\hat{N}_{0}$ is the normalization constant. This equation means that the ground state of $^{16}$O with the $(0s)^4(0p)^{12}$ configuration inherently has a $4\alpha$-cluster degree of freedom. The relative part ({or} second term) of the monopole operator in Eq.~(\ref{eq:e0_4alpha}), $\sum_{k=1}^{4} {4(\vc{R}_{\alpha_k}-\vc{R}_{\rm cm})^2} = {\sum_{k=1}^{3} \mu_{k}{\vc{\xi}_{k}}^{2}}$, can excite the relative motion among the $4\alpha$ particles. In other words, the monopole operator has an ability to populate democratically $4$ $\alpha$-particles by $2\hbar\omega$ with respect to the c.o.m.~coordinate of $^{16}$O. The resultant state, thus, has some amount of the overlap with the $4\alpha$-gas-like state, i.e.~$0^{+}_{6}$, with the $4\alpha$-condensate-like structure~\cite{funaki08}. The overlap value corresponds certainly {to} the monopole matrix element, $M({\rm E0})$. As shown in Ref.~\cite{funaki10}, this $0^{+}_{6}$ state can well {be} described by {a} $4\alpha$-condensate-type microscopic wave function, called the THSR wave function~\cite{tohsaki01}. In this THSR framework, the monopole matrix element to the $4\alpha$-condensate-like state is estimated to be $M({\rm E0})=1.2$~fm$^2$, similar to that in the $4\alpha$ OCM, $M({\rm E0}; {0^+_1}-{0^+_6})=1.0$~fm$^2$ in Table~\ref{tab:4a_ocm}, which is calculated with the use of Eqs.~(\ref{eq:me_e0_ocm}) and (\ref{eq:me_e0_ocm_jacobi}). Thus, the evaluation of the monopole matrix elements using Eqs.~(\ref{eq:me_e0_ocm}) and (\ref{eq:me_e0_ocm_jacobi}) in the OCM framework is useful and gives {a} reasonable {estimate} for the monopole transition to the $\alpha$-$^{12}$C cluster states and $4\alpha$-gas-like states.
 
The mechanism that the $4\alpha$-gas-like state is populated by the monopole transition has a close connection with the mechanism that the Hoyle state with the $3\alpha$-gas-like structure is excited by the monopole transition, in spite of the fact that the ground state of $^{12}$C has {a} shell-model-like compact structure with the main configuration of SU(3) $(\lambda,\mu)=(0,4)$. It is noted that the dominance of the SU(3) symmetry of the $^{12}$C ground state~\cite{elliott58} was confirmed by the {no}-core shell model~\cite{dytrych07}. This SU(3) $(\lambda,\mu)=(0,4)$ wave function is mathematically equivalent to a single $3\alpha$ cluster wave function, according to the Bayman-Bohr theorem, as follow,
\begin{eqnarray}
{\phi_{J=0}({^{12}}{\rm C})} &=& {|(0s)^{4}(0p)^{8};(\lambda,\mu)=(0,4),J^{\pi}=0^{+}\rangle}_{\rm internal} \nonumber \\
&=& \tilde{N}_{0} \sqrt{\frac{4!4!4!}{12!}}\mathcal{A}\left\{ \left[ u_{40}(\vc{\xi}_{2},\frac{8}{3}\nu) u_{40}(\vc{\xi}_{1},2\nu) \right]_{J=0} \phi(\alpha_1) \phi(\alpha_2) \phi(\alpha_3) \right\},\label{eq:su3_12O_3a}
\end{eqnarray}
where $\tilde{N}_{0}$ is the normalization factor. One can see that there exists a similarity between Eqs.~(\ref{eq:su3_16O_4a}) and (\ref{eq:su3_12O_3a}). The latter equation means that the ground state of $^{12}$C originally possesses a $3\alpha$ clustering degree of freedom. As demonstrated in detail in Ref.~\cite{yamada08}, this $3\alpha$ clustering degree of freedom in the $^{12}$C ground state plays an important role in exciting the Hoyle state with the $3\alpha$-gas-like structure by the monopole transition from the shell-model-like ground state. The same story as in the case of the Hoyle state is also realized in the case {of} the $0^+_6$ state with the $4\alpha$-gas-like structure in $^{16}$O {which} is excited by the monopole transition from the ground state with {a} shell-model-like structure, as discussed above.
 
As for the $0^+_5$ state, its main configuration is $\alpha$+$^{12}$C($1^-_1$) with the $P$-wave orbiting of {an} $\alpha$ cluster around the $^{12}$C($1^-_1$) core as mentioned above. According to the Bayman-Bohr theorem, the SU(3) $(\lambda,\mu)=(0,0)$ state of $^{16}$O has no component of the $\alpha$+$^{12}$C($1^-_1$) channel. However, the monopole strength to the $0^+_5$ state is as large as $3$~fm$^2$ (see Table~\ref{tab:4a_ocm}). {This is the reason that} the $0^+_5$ state has small but important components of the $\alpha$+$^{12}$C($0^+_1$,$2^+_1$) and $\alpha$+$^{12}$C($0^+_2$) configurations, as one can see from the {spectroscopic factors} shown in Fig.~\ref{fig:s2_factors_0+_states_16O}. It is noted that the $\alpha$+$^{12}$C($0^+_2$) configuration is likely to be an alternative of the $4\alpha$-gas-like state with the dominant $(0S)^{4}_{\alpha}$ configuration. Since these three configurations, $\alpha$+$^{12}$C($0^+_1$,$2^+_1$) and $(0S)^{4}_{\alpha}$, can be excited from the ground state of $^{16}$O by the monopole operator as discussed above, their respective contributions are coherently added to provide the relatively large monopole strength to the $0^+_5$ state. On the other hand, the {situation of} the $0^+_4$ state, characterized mainly by the higher nodal $\alpha$+$^{12}$C($0^+_1$) state, is similar to the case of the $0^+_5$ state. From Fig.~\ref{fig:s2_factors_0+_states_16O}, the $0^+_4$ state has also small but {non negligible} components of the $\alpha$+$^{12}$C($0^+_1$,$2^+_1$) and $\alpha$+$^{12}$C($0^+_2$) configurations, which contribute to the monopole strength for the $0^+_4$ state.
 
\begin{table}[t]
\begin{center}
\caption{Total energies ($E$) and excitation energies ($E_{x}$) of the $0^+_{1}$ and $0^+_{2}$ states in $^{16}$O as well as those of the $1/2^-_1$ and $3/2^-_1$ states in $^{15}$O obtained by the shell model calculation (see text). We also show r.m.s.~radii ($R$) and {neutron separation energies ($S_n$)} of the $^{16}$O ground state. $E$, $E_x$, and $S_n$ ($R$) are given in unit of MeV (fm).}
\label{tab:shell_model}
\begin{tabular}{cccccccccc}
\hline\hline
\hspace{7mm}    & \multicolumn{3}{c}{Case A} & \multicolumn{3}{c}{Case B} & \multicolumn{3}{c}{Experiment} \\
     &    $E$    &   $E_{x}$  &  $R$ & $E$    &   $E_{x}$  & $R$ & $E$    &   $E_{x}$ & $R$ \\
\hline
$^{16}$O($0^+_1$) & \hspace{2mm}$-127.57$\hspace{2mm} & \hspace{2mm}$ 0.00$\hspace{2mm}  & \hspace{2mm}$ 2.50$\hspace{2mm}  &\hspace{2mm}$-126.79$\hspace{2mm}  & \hspace{2mm}$0.00$\hspace{2mm}  &  \hspace{2mm}$ 2.94$\hspace{2mm}  &\hspace{2mm}$-127.62$\hspace{2mm}  & \hspace{2mm}$0.00$\hspace{2mm} & $2.71$ \\
$^{16}$O($0^+_2$) & \hspace{2mm}$-97.10$\hspace{2mm} & \hspace{2mm}$ 30.43$\hspace{2mm}  &  & \hspace{2mm}$-98.67$\hspace{2mm}  & \hspace{2mm}$28.12$\hspace{2mm}  & &  \hspace{2mm}$-121.57$\hspace{2mm}  & \hspace{2mm}$6.05$\hspace{2mm} &  \\
\hline
$^{15}$O($1/2^-_1$) & \hspace{2mm}$-109.27$\hspace{2mm} & \hspace{2mm}$ 0.00$\hspace{2mm}  &  & \hspace{2mm}$-110.05$\hspace{2mm}  & \hspace{2mm}$0.00$\hspace{2mm}  & & \hspace{2mm}$-111.96$\hspace{2mm}  & \hspace{2mm}$0.00$\hspace{2mm} & \\
$^{15}$O($3/2^-_1$) & \hspace{2mm}$-103.18$\hspace{2mm} & \hspace{2mm}$ 6.09$\hspace{2mm}  & & \hspace{2mm}$-103.77$\hspace{2mm}  & \hspace{2mm}$6.28$\hspace{2mm}  & & \hspace{2mm}$-105.78$\hspace{2mm}  & \hspace{2mm}$6.18$\hspace{2mm}  & \\
\hline
$S_n$            &   $18.30$  & $-$ & $-$ & $16.74$ & $-$ & $-$ & $15.66$ & $-$ & $-$ \\
\hline\hline
\end{tabular}
\end{center}
\end{table}
 
\subsection{Shell model calculation}
 
Table~\ref{tab:shell_model} shows the results of the shell model calculation using the spherical basis (see Sec.~\ref{sub:shell_model}). The binding energies of the ground state of $^{16}$O and those of the $1/2^-_1$ and $3/2^-_1$ states of $^{15}$O are rather well reproduced in the cases A and B. We found that the excitation energy of the $0^+_2$ state is as large as $30$~MeV in the present shell-model calculation, while the experimental value is as small as $6.7$~MeV, the value of which is reproduced by the cluster model (see Table~\ref{tab:4a_ocm}). It is noted that the present shell-model space includes only the $1p1h$ and $2p2h$ configurations up to the $0f1p$ shell. Since the $1p1h$ configurations, in particular, $(1s_{1/2})(0s_{1/2})^{-1}$, $(1p_{3/2})(0p_{3/2})^{-1}$, and $(1p_{1/2})(0p_{1/2})^{-1}$, should give a crucial contribution to the monopole transition strengths, it is instructive to investigate how {strongly} the {components} of the three $1p1h$ configurations, $P(1p1h)$, distribute {among the various} $0^+$ states. We found three energy regions in which the {components} $P(1p1h)$ become significantly large: 1)~$E_x \sim 32$~ MeV with $P(1p1h)\sim 81$~\%, contributed by the $0^+_2$ state, 2)~$E_x \sim 48$ MeV with $P(1p1h)\sim 65$~\%, and 3)~$E_x \sim 60$ MeV with $P(1p1h)\sim 40$~\%. This split of the $1p1h$ component into the three energy region is consistent with the RRPA result~\cite{ma97,lui01} in which the monopole strength concentrates mainly {into} three energy regions (see Fig.~\ref{fig:1}). {This is in line} with the RPA and QRPA calculations (Fig.~3 in Ref.~\cite{gambacurta10}), although the excitation energies of the three energy regions are different. The fact that the excitation energies of the three energy regions in the present shell-model calculations are higher than those in the RRPA, RPA, and QRPA calculations is reasonable, because our calculation employs the spherical harmonic oscillator basis, and the model space taken in its calculation covers only the $1p1h$ and $2p2h$ configurations within the $0s$, $0p$, $0d1s$, and $0f1p$ shells (see Sec.~\ref{sub:shell_model}).

The fact that the present spherical shell model calculation has great difficulty to reproduce the low excitation energies of the $0^+_2$ and $0^+_3$ states is likely to be a common feature of the no-core shell model calculations~\cite{dytrych07}, the FMD calculations~\cite{paar06,papakonstantinou07,papakonstantinou09}, and the coupled-cluster calculations~\cite{wloch05}. {Exceptions are} a few conventional shell-model works, for example, by Brown and Green~\cite{brown66} in 1966 and Arima et al.~\cite{arima67} in 1967, as far as the present authors know. Here it is instructive to briefly present their main results.
 
Brown and Green discussed the low-lying three $0^+$ states of $^{16}$O with the deformed-shell model~\cite{brown66}. It is proposed that the three $0^+$ states ($0^+_1$, $0^+_2$, and $0^+_3$) can be described by the mixture among the $0p0h$, $2p2h$, and $4p4h$ states. In their calculation, the unperturbed energies of the $0p0h$, $2p2h$ and $4p4h$ states are treated as free parameters adjusted to give the observed spectra, although the coupling strengths among the $0p0h$, $2p2h$, and $4p4h$ states are estimated with some approximations based on SU(3) algebra~\cite{brown_contents}. Then they found that the $0^+_1$ state has a dominant configuration of $0p0h$-type, while the main configuration of the $0^+_2$ ($0^+_3$) state is of the $4p4h$-type ($2p2h$-type). With use of the Brown-Green wave functions, Bertsch~\cite{bertsch66} calculated the ${\rm E0}$ transition strength between the $0^+_2$ and $0^+_1$ states, $M({\rm E0};{0^+_2}-{0^+_1})$, defined in Eq.~(\ref{eq:me0_general}). His result is $M({\rm E0})=1.6-2.8$~fm$^2$, corresponding to the experimental data ($3.8$~fm$^2$), although the monopole transition strength between the $0^+_3$ and $0^+_1$ states was not discussed in that paper. The important point in the Brown-Green calculation is that the unperturbed energy of the $4p4h$ configuration is taken to be lower than that of the $2p2h$ one. It is found that if the $2p2h$ state lies lower than the $4p4h$ state, the calculated $B({\rm E2})$ transition rates between the resulting levels for the $0^+$ states and $2^+$ states are difficult to reconcile with the experimental data, although they could not present the reason why the unperturbed energy of the $4p4h$ configuration becomes lower than that of the $2p2h$ one.  
This schematic model proposed by Brown and Green was confirmed by the large-basis spherical shell model calculations mixing the $(0+2+4)\hbar\omega$ excitations~\cite{haxton90,warburton92}. Although they succeed in reproducing the low-lying spectrum of $^{16}$O, the single particle energies are adjusted to fit six low-lying $T=0$ states in $^{16}$O including the $0^+_2$ and $0^+_3$ states~\cite{haxton90}. Thus, the problem of why the excitation energy of the $0^+_2$ state is as small as 6.05~MeV remains unclear in those shell model calculations. 
 
After Brown-Green's work, Arima et al.~proposed a weak coupling picture~\cite{arima67} and showed that one can understand the appearance of the low-lying $0^+_2$ and $0^+_3$ states in $^{16}$O if one
assumes weak coupling between four particles ($4p$) in the $sd$ shell and $4$ hole ($4h$) in $p$ shell. They estimated the coupling strength with a shell model by employing the experimental excitation energy ($E_{x}=0.86$~MeV) of the $1/2^{-}_{1}$ state in $^{19}$F which is described by the $4p$ in $sd$ shell and $1h$ in $p$ shell. They eventually found that the coupling strength between $4p$ and $4h$ in $^{16}$O becomes significantly weak. This weakness of the $4p$ and $4h$ interactions is nothing but the basic assumption of $\alpha$-cluster model in $^{16}$O, i.e.~one can easily imagine that the $4p$ ($4h$) state corresponds to the $\alpha$ ($^{12}$C) cluster. The $4p$ in $sd$ shell can obtain an extraordinarily large binding energy due to the $\alpha$-cluster correlation, and thus the states consisting of $4p$ (in $sd$ shell) and $4h$ (in $p$ shell) can be expected to lie at lower excitation energies, compared with the $2p2h$ states.
 
A structure study of $^{16}$O which explicitly treats the $\alpha$-cluster degree of freedom was performed by Suzuki in 1976 with the semi-microscopic cluster model, $\alpha$+$^{12}$C OCM~\cite{suzuki76}. In this model the relative motion between the $\alpha$ and $^{12}$C clusters is solved, taking into account the coupling between the $\alpha$$-$${^{12}{\rm C}}$ relative motion and the internal rotational motion of $^{12}$C($0^+_1,2^+_1,4^+_1$). Almost all levels of $^{16}$O up to about $E_x \simeq 14$~MeV including the ground state with the dominant configuration of the doubly closed shell structure and the electro-magnetic transition rates (E0, E2, and E3) among them are reproduced well, together with the low-energy ${\alpha}-{^{12}}$C scattering cross sections~\cite{suzuki78} (see Table~\ref{tab:4a_ocm} for the monopole strengths). It is found that a lot of states in $^{16}$O up to $E_x \simeq 14$~MeV have a weak coupling structure of $\alpha$+$^{12}$C, i.e.~loosely bound $\alpha$+$^{12}$C cluster structure. In particular the $0^{+}_{2}$ [$0^{+}_{3}$] state has a weekly coupling structure of the $\alpha$ and $^{12}$C($0^{+}_{1}$) clusters [$\alpha$ and $^{12}$C($2^{+}_{1}$)], in which the $\alpha$ cluster moves predominantly around the $^{12}$C cluster with $S$-wave [$D$-wave] (see also Sec.~\ref{subsec:4a_OCM}). These successes in the $\alpha$+$^{12}$C OCM mean that~1)~the weak coupling picture~\cite{arima67} is realized in $^{16}$O, and the $4p$ ($4h$) state of the $4p4h$ configuration in shell model picture can be interpreted as the $\alpha$ ($^{12}$C) cluster, and 2)~the reason why the energy of the $4p4h$ configurations is lower that of the $2p2h$ one is considered to be the $\alpha$-cluster correlation for the $4p$ state. Although the $\alpha$-cluster structures for the $0^+_2$ and $0^+_3$ states are much different from what Brown and Green thought~\cite{brown66}, one should mention that their conjecture, the energy of the $4p4h$ configurations being lower that of the $2p2h$ one, is remarkable, because it has played an important role in the progress of our understanding the structure of $^{16}$O.

\subsection{Two features of isoscalar monopole transitions}
 
As discussed in Sec.~\ref{subsec:4a_OCM}, the fine structures observed in the experimental monopole strength function in the low energy region up to $E_{x} \simeq 16$ MeV can be rather well reproduced within the $4\alpha$ OCM framework. However, these fine structures are difficult to be reproduced by any mean field theory. This result means that the $\alpha$ clustering degree of freedom is inevitable to reconcile the low-energy behavior of the monopole strength function {in $^{16}$O with experiment}. On the contrary, the RPA calculations look likely to reproduce approximately the three bump structures in the experimental monopole strength function in the higher energy region of $16 \lesssim  E_{x} \lesssim 40$~MeV, although some {normalization} and {energy-shifting} procedures for the calculated strength function is needed to fit the experimental data (see Sec.~\ref{introduction}).
 
Here we should remind that the ground state of $^{16}$O is described {dominantly} by a doubly closed shell structure, $(0s)^{4}(0p)^{12}$, in the $4\alpha$ OCM calculation as well as the RPA, QRPA, and RRPA calculations. As discussed in Sec.~\ref{subsec:4a_OCM}, the doubly closed shell model wave function is mathematically equivalent to a single $\alpha$-cluster wave function. This result means that the ground-state wave function originally has an $\alpha$-clustering degree of freedom together with the single-particle degree of freedom. {In other words, the ground-state wave function of $^{16}$O has a duality of $\alpha$-clustering character and mean-field-type character}.
 
From these facts, one can notice that there exist two types of the isoscalar monopole excitation of $^{16}$O, i.e.~the monopole {excitations} to cluster states {are} dominant in the lower energy part ($E_{x} \lesssim 16$ MeV) of the monopole strength function, whereas the monopole excitation of the one-particle one-hole ($1p1h$) type contributes to the higher energy region ($16 \lesssim E_{x} \lesssim 40$ MeV). {This also is in line with the first $0^+$ excited state of the $\alpha$-particle which is situated at $\sim20$~MeV.} Thus, one can expect that the reproduction of the experimental isoscalar monopole strength function of $^{16}$O in the full energy region up to $E_{x} \sim 40$~MeV will definitely fail, if one does not take into account simultaneously the $\alpha$-cluster-type four-body correlations as well as the $1p1h$- and $2p2h$-type correlations in the structure study of $^{16}$O. In order to tackle the issue, a structure calculation is desirable to be performed in which one uses a huge model space covering fully the $\alpha$-type correlations together with the $1p1h$- and $2p2h$-type correlations
 
We here report on a trial calculation using the $\alpha$+$^{12}$C cluster basis and collective basis for studying {the} isoscalar monopole strength of $^{16}$O~\cite{suzuki89}. {In the latter respect} the symplectic group Sp(6,$R$)~\cite{rosensteel77,rowe85} as the collective basis {was used}. Since the generators of the Sp(6,$R$) group contain the monopole and quadrupole operators with respect to the nucleon coordinates and {conjugate} momenta, {the} group is expected to reproduce the EWSR for the operators. Although there is no effective $NN$ interactions which is suited for cluster-Sp(6,$R$) mixed basis calculation, they took a phenomenological treatment, since their main purpose was to investigate the effect of the Sp(6,$R$) group to the cluster states~\cite{hecht77,hecht78}. They obtained the following two results:~First {comes} that the effect of the {Sp}(6,$R$) basis to the cluster states in the low energy region {is} not very large. It is noted that the $0^+_2$ and $0^+_3$ states (see Fig.~\ref{fig:2}) are the cluster states which are reproduced by the $\alpha$+$^{12}$C model. The second result is that there are three states in the higher energy region ($20 \lesssim E_{x} \lesssim 40$ MeV) which correspond to the three peaks of the monopole excitation function and share about $70~\%$ of the EWSR, although the excitation energies of the three states are higher by about $3$ MeV.  These results support our {finding that} the isoscalar monopole excitations in light nuclei, as mentioned above, {are dominated by two features:~$\alpha$-cluster states at low energies and shell model states at higher energies}. An analysis with {a mixed} $4\alpha$ cluster and symplectic {group} basis may be useful {for this study}.

\section{Summary}\label{sec:summary}
 
We have investigated the monopole strength function {in} the low energy region up to $E_{x} \simeq 16$ MeV within the framework of $4\alpha$ OCM, which has succeeded in {reproducing all} the six $0^+$ states observed up to $E_{x} \simeq 16$ MeV and {in} giving good agreement with all of the available data such as the decay widths, monopole transition strengths, and r.m.s.~radius of the ground state. It was found that the fine structures at the low energy region up to $E_{x} \simeq 16$ MeV in the experimental monopole strength function obtained by the $^{16}$O$(\alpha,\alpha^{\prime})$ experiment is rather satisfactorily reproduced within the $4\alpha$ OCM framework. {On the contrary, mean-field calculations have encountered difficulties} to reproduce the fine structures of the monopole strength function at the low energy region as well as the monopole matrix elements for the $0^{+}_{2}$ ($E_{x}=6.05$~MeV), $0^{+}_{3}$ ($E_{x}=12.05$~MeV), and $0^{+}_{5}$ ($E_{x}=14.01$~MeV) states obtained in the $^{16}$O$(e,e')$ experiment. These results mean that the $\alpha$ clustering degree of freedom is inevitably necessary to reconcile the monopole {strength} (amounting to about $20~\%$ of the EWSR) in the low energy region with {experiment}. On the other hand, the $4\alpha$ cluster model has {difficulties} to reproduce the gross three bump structures of the monopole strength function at the higher energy region of $16 \lesssim  E_{x} \lesssim 40$~MeV. The gross bump structures look likely to be qualitatively reproduced by the mean-field-theory calculations such as RRPA, RPA, and QRPA, although the energy positions of the three bumps and the absolute values of the strength functions quantitatively deviate from the experimental data. In general one can expect the interplay between clustering degrees of freedom and mean-field degrees of freedom. Since the interplay affects the monopole strength function, there is a possibility that the isoscalar monopole strength at the higher energy region of $16 \lesssim E_x \lesssim 40$, in particular at its lower energy part, contains the influence by clustering degrees of freedom. The fact that no mean-field-theory calculations have satisfactorily reproduced the isoscalar monopole strength function even at the higher energy region ($16 \lesssim  E_{x} \lesssim 40$~MeV) 
demonstrates the need to investigate the interplay between clustering and mean-field degrees of freedom in the isoscalar monopole strength function.    
 
From the above results, one {concludes} that there {exist} two features of the isoscalar monopole excitations of $^{16}$O, i.e.~the monopole excitation to cluster states dominates the low energy region ($E_{x} \lesssim 16$~MeV), sharing about $20~\%$ of the EWSR, while that to the $1p1h$-type states looks likely to be predominant at the higher energy region. We indicated that the existence of {these two} types of the monopole {excitations stems} from the fact that the ground state of $^{16}$O with the dominant doubly closed shell configuration $(0s)^{4}(0p)^{12}$ {that is} the dominant SU(3) $(\lambda\mu)=(00)$ symmetry has in fact a dual feature inasmuch as it can equivalently be described by a cluster wave function of the $\alpha$-type as can be shown with the Bayman-Bohr theorem. When the monopole operator activates  $\alpha$-type {degrees} of freedom in the ground state, $\alpha$-cluster states are excited, while in the case of the monopole operator {acting on} the $1p1h$-type degree of freedom in the ground state, collective states of the $1p1h$-type are populated. Thus, one will fail to reproduce the experimental isoscalar monopole strength function of $^{16}$O up to $E_x \sim4 0$~MeV if the $\alpha$-cluster-type four-body correlations as well as the $1p1h$- and $2p2h$-type correlations are not simultaneously taken into account in the structure study of $^{16}$O. 
 
The existence of two features of isoscalar monopole excitation which originates from the dual nature of the ground state seems to be general in light nuclei, and the case of $^{16}$O discussed in the present paper is typical. This is due to {the nuclear SU(3) symmetry which is well verified} in the ground state of light nuclei. According to the Bayman-Bohr theorem, an SU(3) wave function is mathematically equivalent to a cluster-model wave function. Thus the ground state which has a dominant SU(3) symmetry is considered to have the dual nature similar to the case of $^{16}$O, which generates the two features in the isoscalar monopole excitation. However, {this feature} will be vanishing with increasing mass number, {and,} eventually only the $1p1h$-type collective motions are strongly excited, maybe, in the mass region beyond the $fp$-shell nuclei. {The reason for this is that the quality} of the nuclear SU(3) symmetry in the ground state of light nuclei is gradually disappearing because of the stronger effect from the spin-orbit forces {in heavier nuclei}. This means that the dual nature of the ground state is also corroding with increasing mass number. It is an intriguing subject to study theoretically and experimentally how these two features are changing with the mass number. Thus, it is {strongly} hoped that systematic experiments of analyzing the existence of these two features of monopole excitations will be performed in near future.
 
\section*{ACKNOWLEDGMENTS}
 
The present authors thank to Prof.~K.~Kat$\bar{\rm o}$ for useful discussions and comments. This work was partially supported by JSPS (Japan Society for the Promotion of Science) Grant-in-Aid for Scientific Research (C) (21540283) and Young Scientists (B) (21740209).
 
\newpage
\section*{APPENDIX:~Cluster Sum Rule of Isoscalar Monopole Transition}
 
In this Appendix, we first discuss the EWSR of the isoscalar monopole transition for ${^AZ}$ nucleus. Then, we discuss the EWSR of the isoscalar monopole transition within the framework of the OCM, called the OCM-EWSR, in the case of the $n\alpha$ OCM and two-cluster OCM (${^{A_1}{Z_1}}$ and ${^{A_2}{Z_2}}$ with $Z_1=A_1/2$ and $Z_2=A_2/2$) for a self-conjugated nucleus $A=4n$. Finally a general formula of the OCM-EWSR value is presented in the case of the $k$-cluster OCM of $^{Z}A$ nucleus ($k=2,3,\cdots$), composed of the $k$ clusters ($^{Z_1}{A_1}$, $^{Z_2}{A_2}$,$\cdots$, $^{Z_k}A_k$).
 
The EWSR of the isoscalar monopole transition~\cite{yamada08} for ${^AZ}$ nucleus [see Eq.~(\ref{eq:EWSR_nucleon}) for $^{16}$O] is given as
\begin{eqnarray}
&&\sum_{n} (E_n-E_1) {\left| { {\langle {0^{+}_{n}} | \sum_{i=1}^{A} (\vc{r}_i - \vc{R}_{\rm cm} )^2 | {0^{+}_{1}} \rangle}} \right|^2} = \frac{2\hbar^2}{m} \times A \times R^{2},\label{eq:EWSR_nucleon_A}\\
&&R^{2} = \frac{1}{A} {\langle  0^{+}_{1} | \sum_{i=1}^{A}  (\vc{r}_i - \vc{R}_{\rm cm} )^2 | 0^{+}_1 \rangle }, \label{eq:rms_gs_A}
\end{eqnarray}
where $R$ represents the r.m.s radius of the ground state, and other notations are self-evident. Here, we assume that the $NN$ interaction has no velocity dependence. The isoscalar monopole operator in Eq.~(\ref{eq:EWSR_nucleon_A}) can be decomposed as
\begin{eqnarray}
\sum_{i=1}^{4n} (\vc{r}_i - \vc{R}_{\rm cm} )^2 &=& {\sum_{k=1}^{n} \sum_{i=1}^{4} ( \vc{r}_{i+4(k-1)} - \vc{R}_{\alpha_k} )^2} + \sum_{k=1}^{n} {4(\vc{R}_{\alpha_k}-\vc{R}_{\rm cm})^2}, \label{eq:e0_nalpha}\\
                                                       &=& {\sum_{ i \in {A_1} } ( \vc{r}_{i} - \vc{R}_{A_1})^2} + {\sum_{i \in {A_2}}  ( \vc{r}_{i} - \vc{R}_{A_2} )^2} + \frac{A_1 A_2}{A_1+A_2}{\vc{\xi}}^2,\label{eq:e0_A_1_A_2}
\end{eqnarray}
where $\vc{R}_{\alpha_k}=(1/4)\sum_{i=1}^{4}\vc{r}_{i+4(k-1)}$ is the c.o.m.~coordinate of the $k$-th $\alpha$ cluster, and $\vc{R}_{A_1}$ ($\vc{R}_{A_2}$) stands for the c.o.m.~coordinate of the ${^{A_1}{Z_1}}$ (${^{A_2}{Z_2}}$) cluster and $\vc{\xi} = \vc{R}_{A_2} - \vc{R}_{A_1}$ denotes the relative coordinate between the two clusters. As discussed in Sec.~\ref{sub:4a_ocm}, the isoscalar monopole operator in the $n\alpha$ OCM {gives non-zero contribution only for} the second term of Eq.~(\ref{eq:e0_nalpha}), and that in the two-cluster OCM {provides with non-zero contribution only for} the third term of Eq.~(\ref{eq:e0_A_1_A_2}).
 
In the $n\alpha$ OCM, the total wave function of $0^{+}$ state is presented as
\begin{eqnarray}
{ |0^{+}\rangle } = \Phi^{(\rm OCM)}(0^{+}) \prod_{k=1}^{n}\phi(\alpha_{k}),
\end{eqnarray}
where $\phi(\alpha_{k})$ is the internal wave function of $k$-th $\alpha$ cluster and $\Phi^{(\rm OCM)}(0^{+})$ stands for the relative wave function among the $n\alpha$ clusters (see Sec.~\ref{sub:4a_ocm}). Then, the EWSR-OCM is evaluated as
\begin{eqnarray}
&& {\sum_{p} (E^{(\rm OCM)}_p-E^{(\rm OCM)}_1) \left| { {\langle 0^{+}_{p}} | \sum_{i=1}^{4n} (\vc{r}_i - \vc{R}_{\rm cm} )^2 | {0^{+}_{1}} \rangle} \right|^2}, \nonumber \\&& = \sum_{p} (E^{(\rm OCM)}_p-E^{(\rm OCM)}_1) \left| { {\langle \Phi^{(\rm OCM)}({0^{+}_{p}}) | \sum_{k=1}^{n} {4(\vc{R}_{\alpha_k}-\vc{R}_{\rm cm})^2} | \Phi^{(\rm OCM)}({0^{+}_{1}}) \rangle}} \right|^2, \nonumber \\ && = \frac{2\hbar^2}{m} { {\langle \Phi^{(\rm OCM)}({0^{+}_{1}}) | \sum_{k=1}^{n} {4(\vc{R}_{\alpha_k}-\vc{R}_{\rm cm})^2} | \Phi^{(\rm OCM)}({0^{+}_{1}}) \rangle}}, \nonumber \\ && = \frac{2\hbar^2}{m} \times A \times ( R^2 - {R(\alpha)^2} ),\label{eq:EWSR_ocm_n_alpha}
\end{eqnarray}
where $\Phi^{(\rm OCM)}({0^{+}_{{p}}})$ and $E^{(\rm OCM)}_{{p}}$ are the eigenwave function and eigenvalue of the {$p$}-th $0^+$ state obtained by solving the $n\alpha$ OCM equation. Here the $\alpha$-$\alpha$ interaction is assumed to be velocity-independent, and {$R(\alpha)$} stands for the r.m.s.~radius of the $\alpha$ cluster. The derivation of Eq.~(\ref{eq:EWSR_ocm_n_alpha}) should be referred to Sec.~\ref{sub:4a_ocm}. Then, the ratio of the OCM-EWSR to the total EWSR in Eq.~(\ref{eq:EWSR_nucleon_A}) is
\begin{equation}
\frac{{\rm OCM\mbox{\small -}EWSR}}{\rm total~EWSR} = 1 - \left(\frac{{R(\alpha)}}{R}\right)^{2}.
\end{equation}  
This result is {common to all $n$ and hence is the same as Eq.~(\ref{eq:ratio_total_OCM_EWSR}) for $n=4$}. As shown in Sec.~\ref{sub:4a_ocm}, the ratio for the $4\alpha$ OCM is {68}~\%. 
 
In the case of the two-cluster OCM, the total wave function of $0^{+}$ state is presented as
\begin{eqnarray}
{ |0^{+}\rangle } = \Phi^{(\rm 2-clus.)}(0^{+}) \phi(^{A_1}{Z_1}) \phi(^{A_2}{Z_2}),
\end{eqnarray}
where $\phi(^{A_1}{Z_1})$ [$\phi(^{A_2}{Z_2})$] is the internal wave function of the $^{A_2}{Z_2}$ cluster ($^{A_2}{Z_2}$) and $\Phi^{(\rm 2clus.)}(0^{+})$ stands for the relative wave function between the two clusters. Then, the EWSR-OCM is presented as
\begin{eqnarray}
&& {\sum_{p} (E^{(\rm 2clus.)}_p-E^{(\rm 2clus.)}_1) \left| {\langle {0^{+}_{p}} | \sum_{i=1}^{4n} (\vc{r}_i - \vc{R}_{\rm cm} )^2 | {0^{+}_{1}} \rangle } \right|^2}, \nonumber \\ && = \sum_{n} (E^{(\rm 2clus.)}_n-E^{(\rm 2clus.)}_1) \left| { {\langle \Phi^{(\rm 2clus.)}({0^{+}_{n}}) | \frac{A_1 A_2}{A_1+A_2}{\vc{\xi}}^2 | \Phi^{(\rm 2clus.)}({0^{+}_{1}}) \rangle}} \right|^2, \nonumber \\ && = \frac{2\hbar^2}{m} { {\langle \Phi^{(\rm 2clus.)}({0^{+}_{1}}) | \frac{A_1 A_2}{A_1+A_2}{\vc{\xi}}^2 | \Phi^{(\rm 2clus.)}({0^{+}_{1}}) \rangle}}, \nonumber \\ && = \frac{2\hbar^2}{m} \times A \times \left[ R^2 - \frac{A_1}{A} R(A_1)^2 - \frac{A_2}{A} R(A_2)^2 \right],\label{eq:EWSR_ocm_A_1_A_2}
\end{eqnarray}
where $\Phi^{(\rm 2clus.)}({0^{+}_{{n}}})$ and $E^{(\rm 2clus)}_{{p}}$ are the eigenwave function and eigenvalue of the {$p$}-th $0^+$ state obtained by solving the two-cluster OCM equation. Here the two-cluster potential is assumed to be velocity-independent, and $R(A_1)$ [$R(A_2)$] stands for the r.m.s.~radius of the ${^{A_1}{Z_1}}$ (${^{A_2}{Z_2}}$) cluster. It is noted that this EWSR-OCM is realized in the case of the coupled-channel OCM, for example, $\alpha$+$^{12}$C($0^+_1,2^+_1,4^+_1)$), where there is no contribution of the internal monopole transitions in {both the $\alpha$ and $^{12}$C clusters}. Then, the ratio of the OCM-EWSR to the total EWSR in Eq.~(\ref{eq:EWSR_nucleon_A}) is
\begin{equation}
\frac{{\rm OCM\mbox{\small -}EWSR(2-cluster)}}{\rm total~EWSR} = 1 - \frac{A_1}{A}\left(\frac{R(A_1)}{R}\right)^{2} - \frac{A_2}{A}\left(\frac{R(A_2)}{R}\right)^{2}.
\end{equation}
This ratio for the $\alpha$+$^{12}$C OCM is {31~\%}, which is about half of that in the $4\alpha$ OCM.
 
{
Finally let us discuss the OCM-EWSR value, in general, in the case of the $k$-cluster OCM of a $^{Z}A$ nucleus ($k=2,3,\cdots$), composed of the $k$ clusters ($^{Z_1}{A_1}$, $^{Z_2}{A_2}$,$\cdots$, $^{Z_k}A_k$). The total wave function of $0^+$ state is given as
\begin{eqnarray}
{ |0^{+}\rangle } = \Phi^{(k{\rm -clus.})}(0^{+}) {\prod_{i=1}^{k}\phi(^{A_i}{Z_i})},
\end{eqnarray}
where $\phi(^{A_i}{Z_i})$ is the internal wave function of the $^{A_i}{Z_i}$ cluster and $ \Phi^{(k{\rm -clus.})}(0^{+})$ stands for the relative wave function among the $k$ clusters. Then, the EWSR-OCM is presented as
\begin{eqnarray}
&& \sum_{p} (E^{(k{\rm -clus.})}_p-E^{(k{\rm -clus.})}_1) \left| {\langle {0^{+}_{p}} | \sum_{i=1}^{4n} (\vc{r}_i - \vc{R}_{\rm cm} )^2 | {0^{+}_{1}} \rangle } \right|^2, \nonumber \\ && = \frac{2\hbar^2}{m} \times A \times \left[ R^2 - \sum_{i=1}^{k} \frac{A_i}{A} R(A_i)^2 \right],\label{eq:EWSR_ocm_general}
\end{eqnarray}
where $R(A_i)$ denotes the r.m.s.~radius of the $^{A_i}Z_i$ cluster, and we assumed no contributions from the internal monopole transitions in the $^{Z_i}{A_i}$ nucleus ($i = 1, 2, \cdots, k$). The proof of Eq.~(\ref{eq:EWSR_ocm_general}) is similar to those of Eqs.~(\ref{eq:EWSR_ocm_n_alpha}) and (\ref{eq:EWSR_ocm_A_1_A_2}). Then, the ratio of the OCM-EWSR to the total EWSR in Eq.~(\ref{eq:EWSR_nucleon_A}) is
\begin{equation}
\frac{{\rm OCM\mbox{\small -}EWSR(}k{\rm -cluster)}}{\rm total~EWSR} = 1 - \sum_{i=1}^{k} \frac{A_i}{A}\left(\frac{R(A_i)}{R}\right)^{2}.
\end{equation}
The present results can be applied to the monopole transition in neutron-rich nuclei. For example, the ratio for the $\alpha+\alpha+n+n$ OCM of $^{10}$Be amounts to be $68~\%$.
}


\end{document}